\documentclass[a4paper,11pt]{article}

\usepackage[top=18truemm,bottom=25truemm,left=25truemm,right=25truemm]{geometry} 
\usepackage{indentfirst} 

\makeatletter 
\def\section{\@startsection {section}{1}{\z@}{-3.5ex plus -1ex minus -.2ex}{2.3 ex plus .2ex}{\LARGE\bf}}
\makeatother

\makeatletter 
\def\subsection{\@startsection {subsection}{1}{\z@}{-3.5ex plus -1ex minus -.2ex}{2.3 ex plus .2ex}{\Large\bf}}
\makeatother

\makeatletter 
\def\subsubsection{\@startsection {subsubsection}{1}{\z@}{-3.5ex plus -1ex minus -.2ex}{2.3 ex plus .2ex}{\large\bf}}
\makeatother

\usepackage{amsmath, amssymb} 
\if
\makeatletter
	
	\@addtoreset{equation}{sections}
\makeatother
\fi

\usepackage[dvipdfmx]{graphicx} 
\usepackage{float} 
\graphicspath{{./figure/}} 
\usepackage[labelsep=period,figurename=Fig. , tablename=Table]{caption}  
\usepackage{subcaption} 

\usepackage{booktabs} 
\usepackage{multirow}  

\usepackage{url}

\usepackage{color} 
\usepackage{ulem} 
\usepackage{comment} 
\def\vector#1{\mbox{\boldmath $#1$}}
\usepackage{ascmac}
\usepackage{fancybox}

\title{Degree distribution of position-dependent
ball-passing networks in football games}

\usepackage{authblk}
\author[1]{Takuma Narizuka\thanks{Corresponding author.\\
{\it Email address}: physicist.t.n@fuji.waseda.jp (T. Narizuka).\\
}}
\author[2]{Ken Yamamoto}
\author[1]{Yoshihiro Yamazaki}
\affil[1]{Department of Physics, School of Advanced Science and Engineering, Waseda University, Shinjuku, Tokyo 169-8555, Japan}
\affil[2]{Department of Physics, Faculty of Science and Engineering, Chuo University, Bunkyo, Tokyo 112-8551, Japan}

\date{}

\begin{document}
	\maketitle
\begin{abstract}
We propose a simple stochastic model describing the position-dependent ball-passing network in football games.
In this network, a player on a certain area in the divided fields is a node, and a pass between two nodes corresponds to an edge.
Our model is characterized by the consecutive choice of a node dependent on its intrinsic fitness.
We derive the explicit expression of the degree distribution, and find that the derived distribution reproduces the real data quit well.
\end{abstract}

\baselineskip 24pt
\section{Introduction}
In the past years, scientific studies on football have attracted  growing interest \cite{John2002}.
It has been suggested that there are some statistical laws in football dynamics, including goal distributions \cite{Malacarne2000, Greenhough2002, Bittner2008}, temporal features of the ball touches \cite{Mendes2007}, self-similarity of the movement of the ball and players \cite{Kijima2014}.
Complex network analysis \cite{Albert2002, Newman2003} is another approach to extract statistical properties from football games.
In particular, a network in ball passing is composed by nodes and edges corresponding to players and passes, respectively.
Several works have proposed the assessment methods of players and teams based on some network measures such as clustering coefficient, betweenness centrality, and PageRank \cite{Duch2010, Javier2013}.
Furthermore, the structural properties and the spatiotemporal patterns of ball-passing networks have been also investigated \cite{Yamamoto2011, Carlos2013}.

\begin{figure}[H]
	\centering
	\includegraphics[width=14cm]{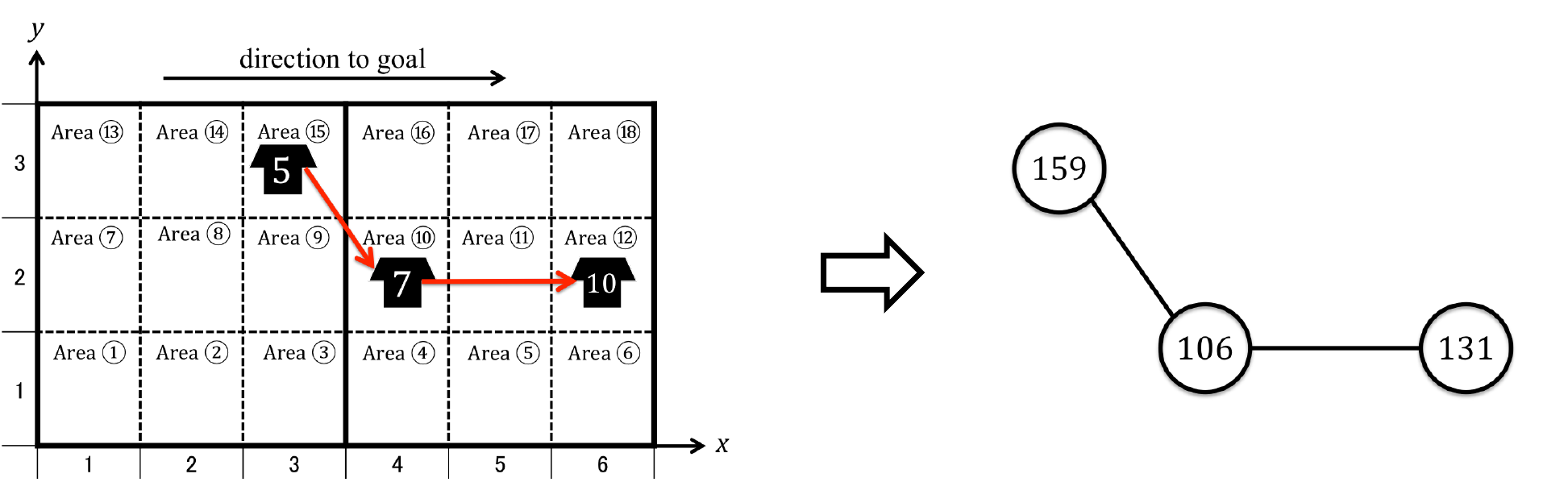}
	\caption{A schematic representation of the creation of a position-dependent ball-passing network. Each player on the three areas in the left figure correspond to the three nodes in the right figure, and the passes between these players correspond to the edges. Each node is given the serial numbers, such as ``159'', ``106'', and ``131'', which are determined from the area and player numbers. Further information about the creation of the network is summarised in our previous paper \cite{Narizuka2014}.}	
	\label{fig:create_net}
\end{figure}

Previously, we have proposed a method to create a ``position-dependent'' ball-passing network in football games \cite{Narizuka2014}.
In this network, a node represents a player on a certain area in the divided fields, and an edge corresponds to a pass between two nodes as shown in Fig. \ref{fig:create_net}.
Note that a node is distinguished by the combination of who makes or receives a pass in which area; that is, one player defines different nodes according to the area of a pass.
The degree $ k $ in this network corresponds to the sum of the numbers of making and receiving passes of each node.
We have obtained the degree distributions of this network from real data.
It was found that they were fitted well by a truncated-gamma distribution, whose probability density function is given by
\begin{align}
	\label{eq:tg}
	f(k) = {\cal N} k^{\nu-1}\mathrm{e}^{-\frac{k}{\lambda}},
\end{align}
where the domain of $ k $ is given as $ 0\le k \le k_{\textrm{max}} $, and $ \nu $, $ \lambda $, and $ k_{\textrm{max}}$ are the fitting parameters.
The normalization constant $ {\cal N} $ depends on $ k_{\textrm{max}}$.
By introducing $ k_{\textrm{max}} $, we obtained the better fitting results than the ordinary gamma distribution.

In our previous paper \cite{Narizuka2014}, we have also proposed the numerical model called the ``Markov-chain model'' describing the ball-passing process.
In this model, the passing sequence is simulated by a Markov chain.
The ball-possession probability $ x_{j}^{(t)} $ of node $ j $ at time $ t $ is calculated by the Markov chain
\begin{align}
	x_{j}^{(t+1)} = \sum_{i=1}^{N} x_{i}^{(t)}P_{i\to j},  \nonumber
\end{align}
where $ P_{i\to j} $ is the transition probability of a ball from node $ i $ to $ j $, and $ N $ is the total number of the nodes.
We have assumed that  $ x_{j}^{(t)} $ is proportional to the degree of node $ j $.
For the football games, $ P_{i\to j} $ is assumed to be defined as
\begin{align}
	P_{i \to j}  = \frac{Q(r_{ij}) \times R(L_{j})}{Z} ,  \nonumber
\end{align}
where $ Q(r_{ij}) $ represents the rate of completing a pass dependent on the distance $ r_{ij} $ between the two nodes $ i $ and $ j $.
$ R(L_{j}) $ denotes the factor for existence probability of the player receiving a pass, where $ L_{j} $ is the distance of the node $ j $ from its home position.
(The home position is assigned for each player randomly.)
$ Z $ is the normalization constant which is determined to satisfy $ \displaystyle{\sum_{j} P_{i\to j}} $ =1.
We have discussed numerically the property of the cumulative distribution $ G(x) $ of the ball-possession probabilities $ x_{j}^{(t)} $, as a substitute for the degree distribution, and compared $ G(x) $ with the truncated-gamma distribution \eqref{eq:tg}.
We have found that $ G(x) $ depends mainly on the factor $ R(L) $, and reproduces the truncated-gamma distribution by choosing an appropriate function for $ R(L) $.
Meanwhile, in the simplified condition where $ P_{i \to j} \propto R(L_{j}) $, we get $ x_{j}^{(t)} = R(L_{j})$.
Then, the probability distribution $ g(x) $ of $ x_{j}^{(t)} $, defined as $ g(x) \equiv -dG(x)/dx $, and the probability density function $ \rho(L) $ of $ L_{j} $ hold the relation, $ g(x) dx = \rho(L) dL$.
And, $ g(x) $ can be expressed as
\begin{align}
	g(x) = \rho\left[R^{-1}(Zx) \right] \left| \frac{d }{d x} \left[R^{-1}(Zx) \right] \right|.
	\label{eq:g(x)}
\end{align}

On the basis of the above results, the present paper derives the explicit expression of the degree distribution based on the above Markov-chain model.
For the derivation, we employ the framework of the fitness model, and extend it to the case where networks contain the temporal feature, i.e., the time series of the passes.

\section{Extended Markov-chain model for ball passing}
\subsection{Setup}
In this section, we propose an extended model based on the above Markov-chain model.
In the extension, we take into account the interaction of two teams $ A $ and $ B $, each of which consists of 11 players.
A node expresses a player on a divided area in the field.
The field is divided into $ 2\Delta $ sections along the goal direction, and $ \Delta $ sections along the  direction vertical to the goal direction.
(The field division shown in Fig. \ref{fig:create_net} is the case $\Delta=3$.)
Thus, the total number of divided areas is $ 2\Delta^{2} $, and each area equally has 22 nodes; nodes are distributed uniformly over all areas.
The number of nodes in each team denoted by $ N_{A} $ and $ N_{B} $, is $ 11\times 2\Delta^{2} $, and the total number of nodes are $ N=N_{A} + N_{B} $.
Each node in team $ A $ and $ B $ is given the serial numbers $ a $ ($ a=1 , \ldots  N_{A}$) and $ b $ ($ b= N_{A}+1 , \ldots  , N_{A}+N_{B} $), respectively.
For each player, we assign one of areas in the divided field as the home position, and define the distance $ L $ between the home position and the position for each node.
$ L $ is the quantity defined for all nodes individually, and we call $ L $ a ``fitness'' hereafter.
(Fitness is a term usually used in complex network analysis \cite{Caldarelli2002, Boguna2003}.)
The probability distribution of $ L $ for each team is expressed as $ \rho_{A}(L) $ and $ \rho_{B}(L)  $.

Now, the passing sequences are assumed to be random transfer of the ball between nodes.
For the transition probability between two nodes, we assume the following four forms depending on the teams to which these nodes belong:
\begin{subequations}
\begin{align}
	P_{a'\to a} &= \eta_{A} R_{A}(L_{a}) / Z_{A} , \label{eq:tr_prob_a}\\
	P_{b'\to b} &= \eta_{B} R_{B}(L_{b}) / Z_{B} ,  \\
	P_{a'\to b} &= (1-\eta_{A}) R_{B}(L_{b}) / Z_{B} , \\
	P_{b'\to a} &= (1-\eta_{B}) R_{A}(L_{a}) / Z_{A} . \label{eq:tr_prob_d}
\end{align}
\end{subequations}
Here $ \eta_{A} $ and $ \eta_{B} $ represent the ball-passing probability within the same teams, and the existence probability $ R(L_{j}) $ of node $ j $ is a monotonically decreasing function of $ L $.
Owing to the normalization of $ P_{i\to j} $ 
\begin{align}
	\sum_{j=1}^{N} P_{a'\to j} = 1, \hspace{0.5cm} \sum_{j=1}^{N} P_{b'\to j} = 1, \nonumber
\end{align}
the coefficients $ Z_{A} $ and $ Z_{B} $ are expressed as
\begin{align}
	Z_{A} = \sum_{a' = 1}^{N_{A}} R_{A}(L_{a'}), \hspace{0.5cm} Z_{B} = \sum_{b' = N_{A}+1}^{N} R_{B}(L_{b'}) . \nonumber
\end{align}
The explicit forms of $ \rho_{A}(L) $ and $ R_{A}(L) $ are discussed in Sec. 3.

\begin{figure}[H]
	\centering
		\includegraphics[width=7cm]{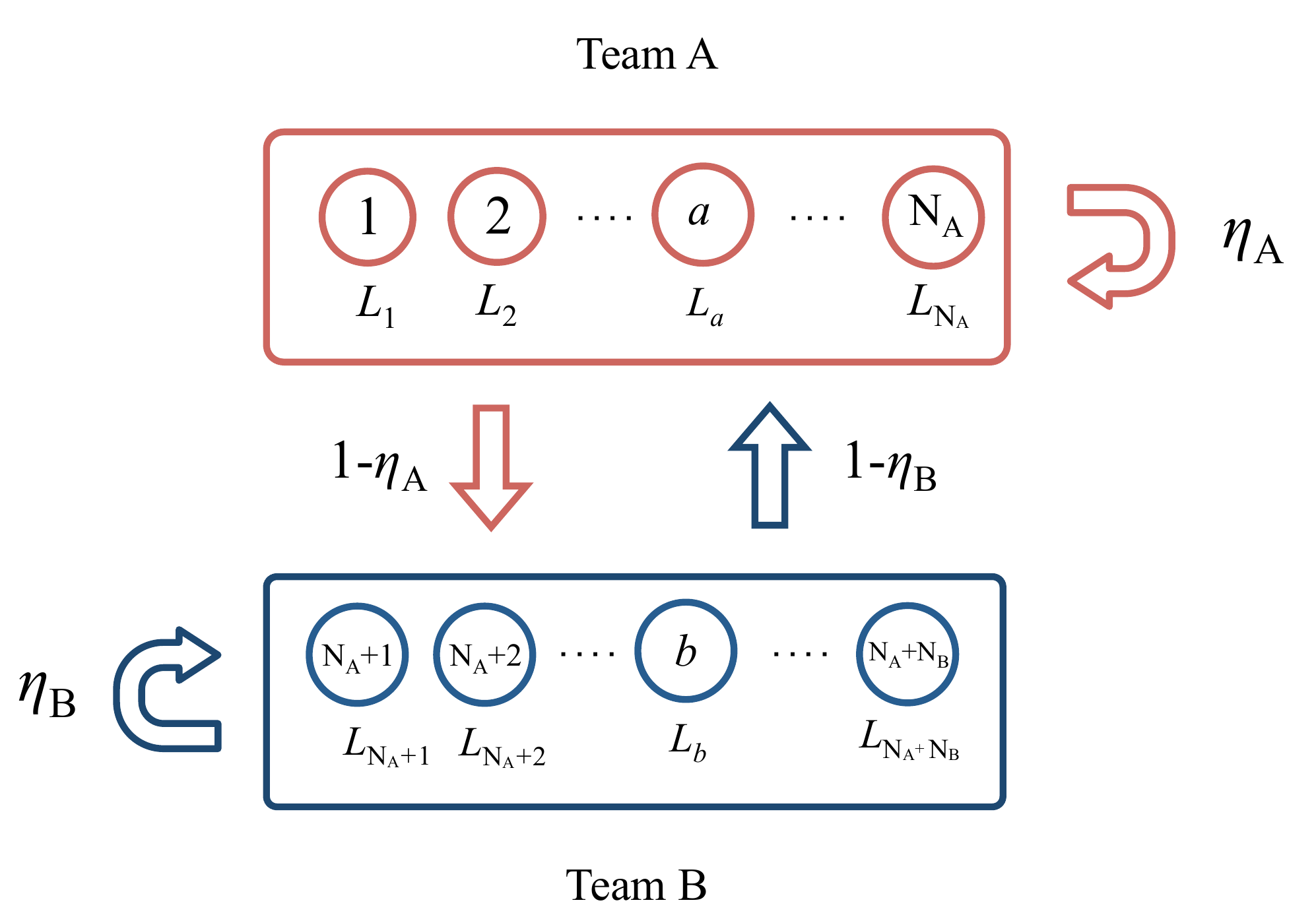}
		\caption{A schematic representation of the extended model. Teams $ A $ and $ B $ have $ N_{A} $ and $ N_{B} $ nodes respectively, and each node has the fitness $L$.}
		\label{fig:tr_diagram}
\end{figure}

\subsection{Ball-possession probability}
We introduce the ball-possession probability $ x_{j}^{(t)} \ (j = 1, \ldots, N_{A}, N_{A}+1, \ldots , N) $ for the node $ j $ at time $ t $, and the probability vector $ \vector{x}^{(t)} $ is defined as
\begin{align}
	\vector{x}^{(t)} &= [ x_{1}^{(t)}, \ldots , x_{a}^{(t)}, \ldots, x_{N_{A}}^{(t)}, x_{N_{A}+1}^{(t)}, \ldots,  x_{b}^{(t)}, \ldots, x_{{N}}^{(t)}] .  \nonumber
\end{align}
$ \vector{x}^{(t)} $ is normalized at each $ t $ as
\begin{align}
	\sum_{a=1}^{N_{A}} x_{a}^{(t)} &= X^{(t)}_{A},  \nonumber\\
	\sum_{b=N_{A}+1}^{N} x_{b}^{(t)} &= X^{(t)}_{B},  \nonumber
\end{align}
where $ X_{A}^{(t)} $ and $ X_{B}^{(t)} $ are the ball-possession probabilities for team $ A $ and $ B $ which satisfy $  X^{(t)}_{A} + X^{(t)}_{B} = 1 $.

The time evolution of $ x_{a}^{(t)} $ is given by the Markov chain
\begin{align}
	 x_{a}^{(t)} &= \sum_{j=1}^{N} x_{j}^{(t-1)} P_{j\to a}.
	 \label{eq:mc_a}
\end{align}
Substituting Eqs. \eqref{eq:tr_prob_a} and \eqref{eq:tr_prob_d} into Eq. \eqref{eq:mc_a}, we obtain 
\begin{align}
	x_{a}^{(t)}      &= \sum_{a'=1}^{N_{A}} x_{a'}^{(t-1)} \eta_{A} \frac{R_{A}(L_{a})}{ Z_{A} }+ \sum_{b'=N_{A}+1}^{N} x_{b'}^{(t-1)} (1-\eta_{B}) \frac{R_{A}(L_{a})}{ Z_{A} }  \nonumber \\
	                   &= \left[(1-\eta_{B})   + (\eta_{A}+\eta_{B}-1)  X^{(t-1)}_{A} \right] \frac{R_{A}(L_{a})}{ Z_{A} } .
	                   \label{eq:bpp1}
\end{align}
By taking summation for $ a $ from $ 1 $ to $ N_{A} $ in the both sides of Eq. \eqref{eq:bpp1}, the following recurrence relation for $ X^{(t)}_{A} $ is obtained:
\begin{align}
	X^{(t)}_{A} &= (1-\eta_{B}) + (\eta_{A}+\eta_{B}-1)X^{(t-1)}_{A}.  \nonumber
\end{align}
The solution is
\begin{align}
	X^{(t)}_{A} &= \frac{1-\eta_{B}}{ 2-\eta_{A}-\eta_{B}} - \frac{(1-\eta_{B})+(\eta_{A}+\eta_{B}-2)X_{A}^{(0)}}{ 2-\eta_{A}-\eta_{B}}(\eta_{A}+\eta_{B}-1)^{t}.
	\label{eq:bppA}
\end{align}
Substituting Eq. \eqref{eq:bppA} into Eq. \eqref{eq:bpp1}, we have
\begin{align}
	x_{a}^{(t)} &= \left[\frac{1-\eta_{B}}{ 2-\eta_{A}-\eta_{B}} 
			 		- \frac{(1-\eta_{B})+(\eta_{A}+\eta_{B}-2)X_{A}^{(0)}}{ 2-\eta_{A}-\eta_{B}}(\eta_{A}+\eta_{B}-1)^{t} \right] \frac{R_{A}(L_{a})}{ Z_{A} }.
          		      \label{eq:bpp2}
\end{align}
Here, the second term of Eq. \eqref{eq:bpp2} decreases exponentially with time $ t $, because $ |\eta_{A} + \eta_{B} -1| < 1 $ holds from $ 0 < \eta_{A}, \ \eta_{B} < 1 $.
The total number $ T $ of ball transitions in one game is more than 1000 (see Table \ref{tb:params_data} in detail), so the exponential term becomes ignorably small compared with the first constant term.
In the stationary state, the ball-possession probability for node $ a $ in team $ A $ can be written as
\begin{align}
	x_{a} = \frac{1-\eta_{B}}{2-\eta_{A}-\eta_{B}}   \ \frac{R_{A}(L_{a})}{ Z_{A} }.  \nonumber
\end{align}
The same discussion is applied to team $ B $.
Therefore, we can write the ball-possession probability for a node with fitness $ L $ in team $ A $ and $ B $ as
\begin{align}
	x_{A}(L) &= \frac{1-\eta_{B}}{2-\eta_{A}-\eta_{B}}    \ \frac{R_{A}(L)}{ Z_{A} },  \nonumber\\
	x_{B}(L) &= \frac{1-\eta_{A}}{2-\eta_{A}-\eta_{B}}    \ \frac{R_{B}(L)}{ Z_{B} }.  \nonumber
\end{align}

\subsection{Derivation of the degree distribution}
Here, we focus on a node $ a $ in team $ A $ whose fitness is $ L $, and derive the in-degree and out-degree distributions $ f_{A}^{(\textrm{in})}(k') $ and $ f_{A}^{(\textrm{out})}(k") $.
The in-degree of node $a$ increases when the node receives a pass from another node in the same team.
This event takes place at each time step with probability
\begin{align}
	\sum_{a'=1}^{N_{A}} x_{a'} P_{a'\to a} 	&= \eta_{A} x_{A}(L).  \nonumber
\end{align}
On the other hand, the out-degree of node $ a $ increases by the event that the node makes a pass to another node in the same team, and it occurs with probability
\begin{align}
	\sum_{a'=1}^{N_{A}} x_{a} P_{a\to a'} = \eta_{A} x_{A}(L).  \nonumber
\end{align}
Here, we introduce the conditional probability $ q_{A}^{(\textrm{in})}(k'|L) $ $ \left( q_{A}^{(\textrm{out} )}(k"|L) \right) $ that a node with fitness $ L $ gets the in-degree $ k' $ (out-degree $ k" $) at time $ T $.
The in-degree or out-degree of the node with fitness $ L $ is increased by one at each time step with the probability $ \eta_{A}x_{A}(L) $.
Since these events are independent each other, $ q_{A}^{(\textrm{in})}(k'|L) $ and $ q_{A}^{(\textrm{out} )}(k"|L) $ are given as the binomial distribution:
\begin{align}
	 q_{A}^{(\textrm{in})}(k'|L) &=   
	\binom{T}{k'}
	 [\eta_{A} x_{A}(L)]^{k'} [1-\eta_{A} x_{A}(L)]^{T-k'} , \nonumber\\[8pt]
	 q_{A}^{(\textrm{out})}(k"|L) &=   
	\binom{T}{k"}
	 [\eta_{A}x_{A}(L)]^{k"} [1-\eta_{A}x_{A}(L)]^{T-k"},  \nonumber
\end{align}
where $ T $ is the total number of ball passing.
If field division $ \Delta $ is large enough, $ L $ can be regarded as the continuous variable, so that we calculate $ f_{A}^{(\textrm{in})}(k') $ and $ f_{A}^{(\textrm{out})}(k") $ as follows:
\begin{align}
	 f_{A}^{(\textrm{in})}(k') 
	 &= 
	\int_{0}^{\infty}
	\binom{T}{k'}
	 [\eta_{A} x_{A}(L)]^{k'} [1-\eta_{A} x_{A}(L)]^{T-k'} \rho_{A}(L) dL,  \nonumber\\[8pt]
	 f_{A}^{(\textrm{out})}(k") 
	&=
	\int_{0}^{\infty}
	\binom{T}{k"}
	 [\eta_{A}x_{A}(L)]^{k"} [1-\eta_{A}x_{A}(L)]^{T-k"} \rho_{A}(L) dL.  \nonumber
\end{align}

Next, we derive the degree distribution $ f_{A}(k) $.
Since the degree $ k $ is the sum of the in-degree $ k' $ and the out-degree $ k" $,
the probability $ q_{A}(k|L) $ that the node with fitness $ L $ has the degree $ k $ is given by the convolution
\begin{align}
q_{A}(k|L)
	&= \sum_{k=k'+k"} q_{A}^{(\textrm{in})}(k'|L)q_{A}^{(\textrm{out})}(k"|L) \nonumber \\
	&= 
	\binom{2T}{k}
	  [\eta_{A}x_{A}(L)]^{k} [1-\eta_{A}x_{A}(L)]^{2T-k} .  \nonumber
\end{align}
Hence, the degree distribution $ f_{A}(k) $ is obtained as follows:
\begin{align}
	 f_{A}(k) 
	 &=
	 \int_{0}^{\infty}q_{A}(k|L)\rho_{A}(L) dL \nonumber \\
	 &= 
	\int_{0}^{\infty}
	\binom{2T}{k}
	 [\eta_{A}x_{A}(L)]^{k} [1-\eta_{A}x_{A}(L)]^{2T-k} \rho_{A}(L) dL  \nonumber\\
	 &= \int_{0}^{\infty}
	\binom{2T}{k}
	 \left[ \frac{R_{A}(L)}{Z_{A}'}\right]^{k} \left[1- \frac{R_{A}(L)}{Z_{A}'} \right]^{2T-k} \rho_{A}(L) dL,
	 \label{eq:fk_1} 
\end{align}
where $ Z_{A}' $ is defined as 
\begin{align}
	Z_{A}' = \frac{2-\eta_{A}-\eta_{B}}{\eta_{A}(1-\eta_{B})} Z_{A}.  \nonumber
\end{align}
By the variable transformation $ u =R_{A}(L)/Z_{A}' $,
$ f_{A}(k)  $ is rewritten as
\begin{align}
	f_{A}(k) 
	= -\int_{u_{\textrm{min}}}^{u_{\textrm{max}}}
	\binom{2T}{k}
	 	[u]^{k} [1-u]^{2T-k}  \rho_{A}\left[ R^{-1}_{A}\left(  Z_{A}'u \right) \right]
	       	  \frac{d}{du} \left[ R^{-1}_{A}\left(Z_{A}' u\right)\right]  du,
	\label{eq:fk_2}
\end{align}
where
\begin{align}
	u_{\textrm{min} } &=  \frac{R_{A}(\infty)}{Z_{A}'} = 0,  \nonumber \\ 
	u_{\textrm{max} } &= \frac{R_{A}(0)}{Z_{A}'}.  \nonumber
\end{align}
From the saddle point method, the following approximation of Eq. \eqref{eq:fk_2} is obtained when $ 2T \gg k $ (see Appendix for derivation):
\begin{align}
	f_{A}(k) 
	&\simeq -\rho_{A}\left[ R^{-1}_{A}\left(  Z_{A}'k/2T \right) \right]
	 	 \frac{d}{dk} \left[ R^{-1}_{A}\left(Z_{A}' k/2T \right)\right].
        \label{eq:fk_sim}
\end{align}

\section{Application to real data}
\begin{figure}[H]
	\begin{minipage}{.5\textwidth}
		\centering
		\includegraphics[width=7.5truecm]{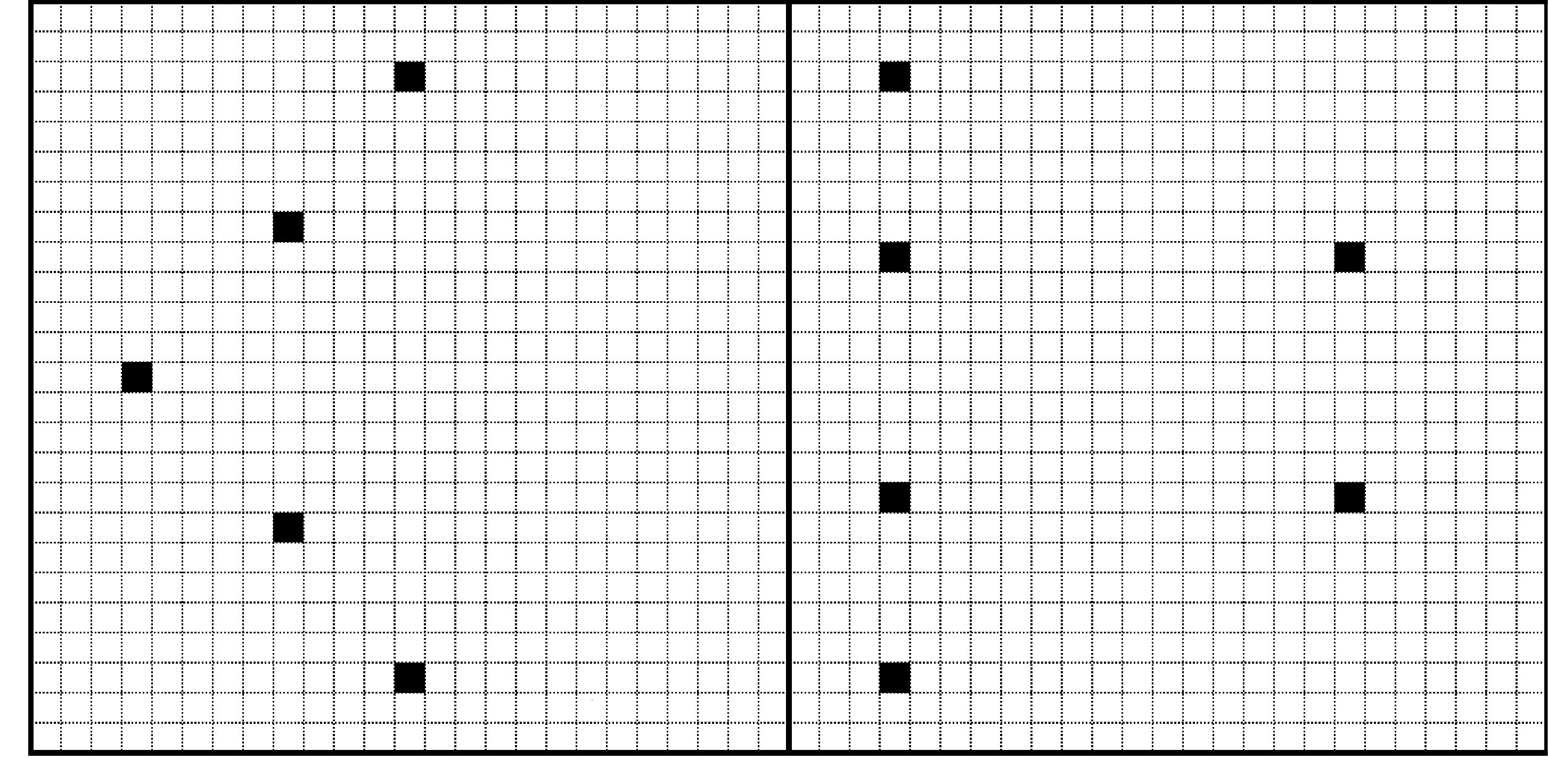}
		\caption*{(i)}
	\end{minipage}
	\begin{minipage}{.5\textwidth}
		\centering
		\includegraphics[width=6.0truecm]{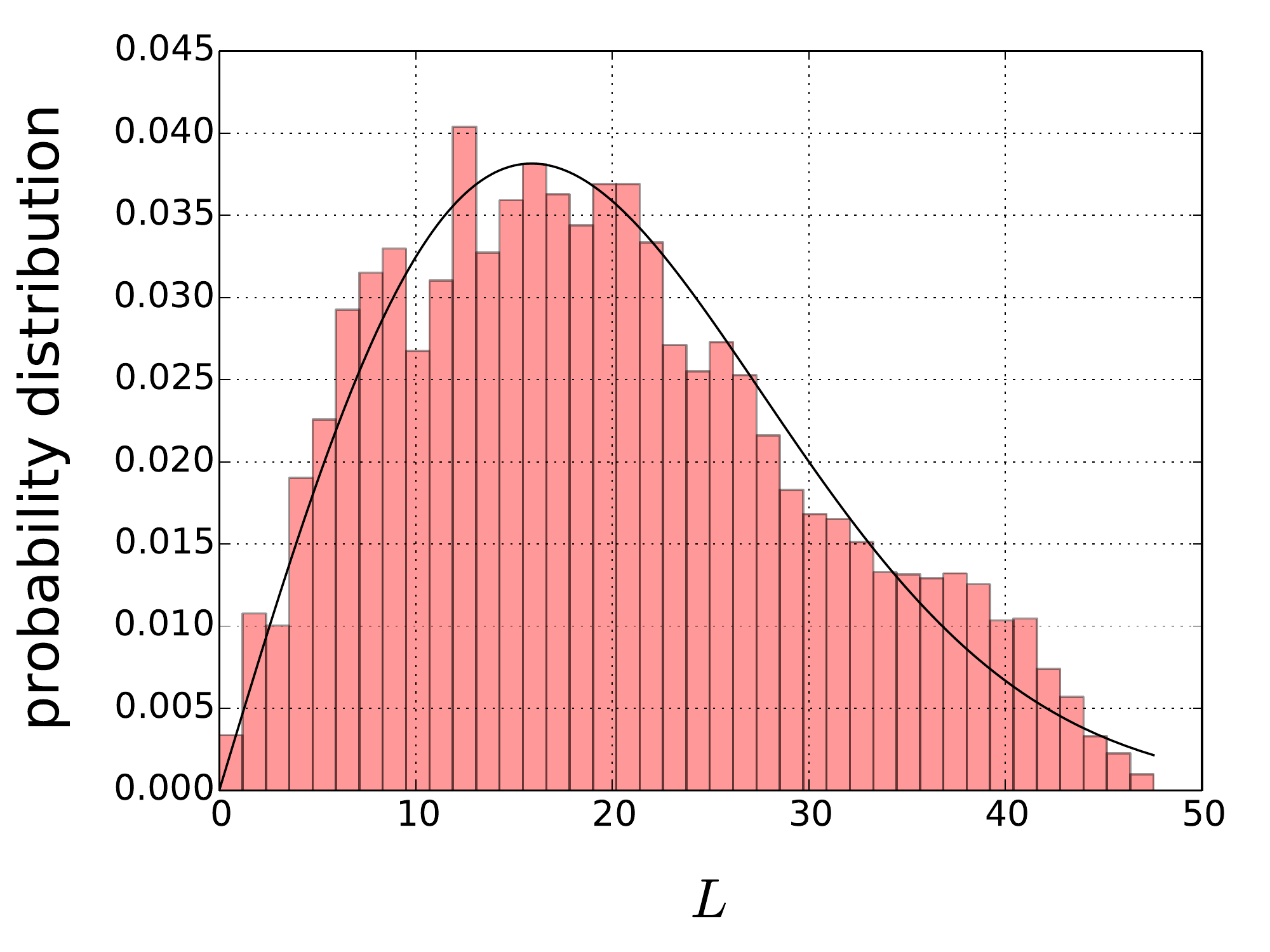}
		\caption*{(ii)}
	\end{minipage}
	\caption{(i) A typical example of the configuration for the home position of each player in the $ 50 \times 25 $ field division. (ii) The probability distribution of $ L $ in this home position. The solid curve is the fitting result by Eq. \eqref{eq:rho}.}
	\label{fig:homeposi}
\end{figure}

To compare our theoretical result with the real data, we have to fix two functions $R_{A}(L)$ and $\rho_{A}(L)$ in Eq. \eqref{eq:fk_2}.
Since each player is less likely to visit areas distant from its home position, $ R_{A}(L) $ corresponding to the existence probability of each player should be a monotonically decreasing function.
Here, we assume
\begin{align}
	R_{A}(L) &\propto \exp[{-\left(L/{\beta}\right)^{\frac{2}{m}}}],
	\label{eq:R}
\end{align}
where $ \beta $ is the scale parameter and $ m $ is the shape parameter.
We adopt this form because it can express some different decreasing functions by changing $ m $; it becomes Gaussian function for $ m=1 $, and exponential function for $ m=2 $.
Next, we determine the function $ \rho_{A}(L) $, which corresponds to the number of nodes whose distance from its home position is $ L $.
Since such a node is on a circle centered at its home position with radius $ L $, and nodes are assumed to be distributed uniformly over all areas, $\rho_{A}(L)$ grows linearly for small $L$, but turns to decrease for large $L$ because of the finite field size.
The following function
\begin{align}
	\rho_{A}(L) = \frac{2}{\omega^{2}} L \exp \left[- \left(L/\omega\right)^{2}\right],
	\label{eq:rho}
\end{align}
satisfies these properties.
The factor $ {\exp}[{-(L/\omega)^{2}}] $ expresses the effect of the field boundary.
We have checked whether this function is good approximation of the probability distribution of $ L $ (see Fig. \ref{fig:homeposi} for a typical example).
And also, we have found that $ \omega $ becomes almost constant for any configurations of the home position as follows.
In the condition where $ \Delta =25 $, we assigned one of areas to each player randomly as the home position, and examined the probability distribution of $ L $ for 1000 samples.
For each sample, we fitted the probability distribution of $ L $ by using Eq. \eqref{eq:rho}, and obtained $ \omega = 22.72 \pm 1.34 $.
This result implies that almost the same value of $ \omega $ is obtained irrespective of the configuration of the home position.

Finally we derive the explicit expression of the degree distribution for the ball-passing network.
Substituting Eqs. \eqref{eq:R} and \eqref{eq:rho} into Eq. \eqref{eq:fk_2}, and introducing the parameter $ \mu = (\omega/\beta)^{2/m} $, we get
\begin{align}
	f_{A}(k) =\int_{u_{\textrm{min}}}^{u_{\textrm{max}}}
	\binom{2T}{k}
	 	[u]^{k} [1-u]^{2T-k}  
		  \frac{m}{ \mu}\left(  \frac{\log\left( u_{\mathrm{max}}/u\right)}{\mu} \right)^{m-1} \frac{1}{u} \ 
						\exp \left[ -\left( \frac{\log\left( u_{\mathrm{max}}/u\right)}{\mu} \right)^{m} \right] du,
	\label{eq:fk_real}
\end{align}
where
\begin{align}
	u_{\textrm{min} } &= 0  \nonumber \\
	u_{\textrm{max} } &= \frac{1}{Z_{A}'}= \frac{1}{Z_{A}} \frac{\eta_{A}(1-\eta_{B})}{2-\eta_{A}-\eta_{B}} . \nonumber
\end{align}

\begin{figure}[H]
	\begin{minipage}{.5\textwidth}
		\centering
		\includegraphics[width=6.5truecm]{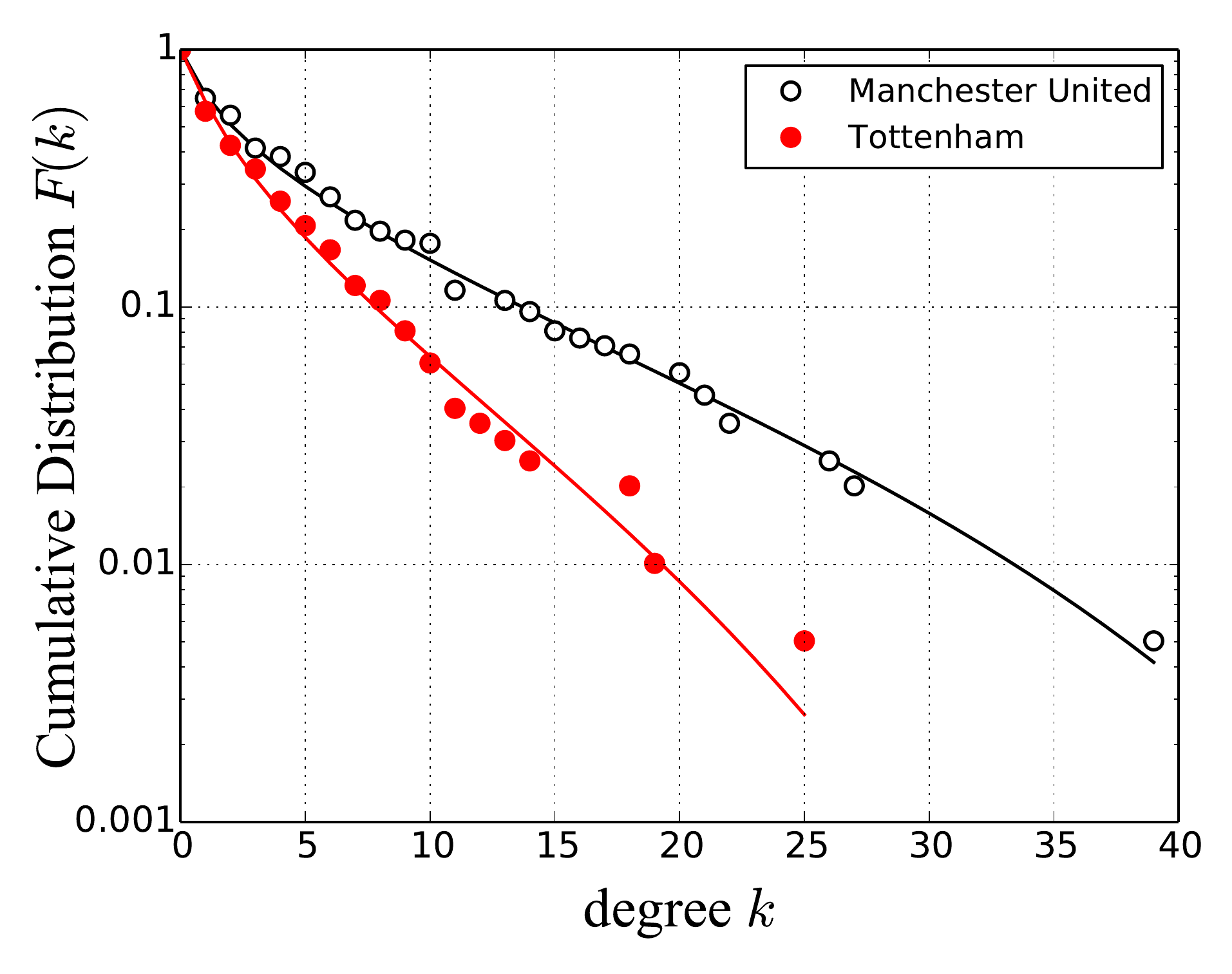}
		\caption*{(ix)}
	\end{minipage}
	\caption{Fitting for the cumulative distributions of real data by using Eq. \eqref{eq:fk_real}. The panels (i)-(ix) correspond to the nine games in Table \ref{tb:game_data}. The solid curves show the cumulative distribution function of Eq. \eqref{eq:fk_real}.}
	\label{fig:degree}
\end{figure}

Now, we analyze the real data.
We have investigated 18 networks obtained from 9 real games as shown in Table \ref{tb:game_data}, with the field division $ \Delta = 3$.
Degree distribution [Eq.\eqref{eq:fk_real}] contains the six parameters, $ \mu $, $ m $, $ T $, $ N_{A} $, $ \eta_{A} $ and $ \eta_{B} $.
The value of $ N_{A} $ is 198 when $ \Delta = 3$.
$ T $ is evaluated by directly counting the number of ball transitions in a whole game, and 
$ \eta_{A} $ and $ \eta_{B} $, the ball-passing probability between the same teams, are obtained as the ratio of the number of ball transitions within the same teams and $ T $.
These values are summarised in Table \ref{tb:params_data}.
Then, we regard the remaining two parameters $ \mu $ and $ m $ as the control parameters for fittings.
Figure \ref{fig:degree} shows the cumulative degree distributions of each network in a single logarithmic scale.
The solid curves in each panel are the cumulative distribution $ F(k) $ of Eq. \eqref{eq:fk_real}.
The values of $ \mu $ and $ m $ for fittings are summarised in Table \ref{tb:params_fit}.
We find that all real data in Fig. \ref{fig:degree} are in good agreement with Eq. \eqref{eq:fk_real}.

\begin{table}[H]
	\centering
 	\caption{Real game data for the analysis.}
	\vspace*{-0.3cm} 
	\label{tb:game_data}
 		\begin{tabular}{cccccc}
		\toprule
		      & Game                           & Place & Date & Score             &  Competition           \\ \toprule
		(i)   & Japan vs Paraguay              & South Africa  & 2010.06.19 & 0-0  & Wcup 2010              \\
		(ii)  & Japan vs Vietnam               & Japan        & 2011.10.07 & 1-0  & Kirin Challenge Cup    \\
		(iii) & Japan vs Tajikistan            & Japan        & 2011.10.11 & 8-0  & WCup Asian qualifier   \\
		(iv)  & Japan vs North Korea           & North Korea  & 2011.11.15 & 0-1  & WCup Asian qualifier   \\ 
		(v)   & Spain vs Italy                 & Poland       & 2012.06.10 & 1-1  & Euro 2012              \\ 
		(vi)  & Germany vs Holland             & Ukraine      & 2012.06.13 & 2-1  & Euro 2012              \\
		(vii) & Mainz vs Hertha                & Germany      & 2013.09.28 & 1-3  & Bundesliga 13-14       \\
		(viii)& Manchester City vs Everton     & England      & 2013.10.05 & 3-1  & Premiere league 13-14  \\
		(ix)  & Manchester United vs Tottenham & England      & 2014.01.01 & 1-2  & Premiere league 13-14  \\               
		\bottomrule 
		\end{tabular}
\end{table}
\begin{table}[H]
	\centering
 	\caption{The parameters obtained from the real data.}
	\vspace*{-0.3cm} 
	\label{tb:params_data}
 		\begin{tabular}{cccccc}
		\toprule
		      & Game                           & $ T $ & $ \eta_{\textrm{A}} $ & $ \eta_{\textrm{B}} $ \\ \toprule
		(i)   & Japan vs Paraguay              & 1457  & 0.51  & 0.58    \\
		(ii)  & Japan vs Vietnam               & 1345  & 0.65  & 0.55    \\
		(iii) & Japan vs Tajikistan            & 1438  & 0.69  & 0.39    \\
		(iv)  & Japan vs North Korea           & 1094  & 0.59  & 0.55    \\ 
		(v)   & Spain vs Italy                 & 1471  & 0.72  & 0.62    \\ 
		(vi)  & Germany vs Holland             & 1390  & 0.65  & 0.69    \\
		(vii) & Mainz vs Hertha                & 1138  & 0.47  & 0.62    \\
		(viii)& Manchester City vs Everton     & 1314  & 0.65  & 0.60    \\
		(ix)  & Manchester United vs Tottenham & 1215  & 0.61  & 0.52    \\               
		\bottomrule 
		\end{tabular}
\end{table}

\clearpage
\begin{figure}[H]
	\begin{minipage}{.5\textwidth}
		\centering
		\includegraphics[width=6.5truecm]{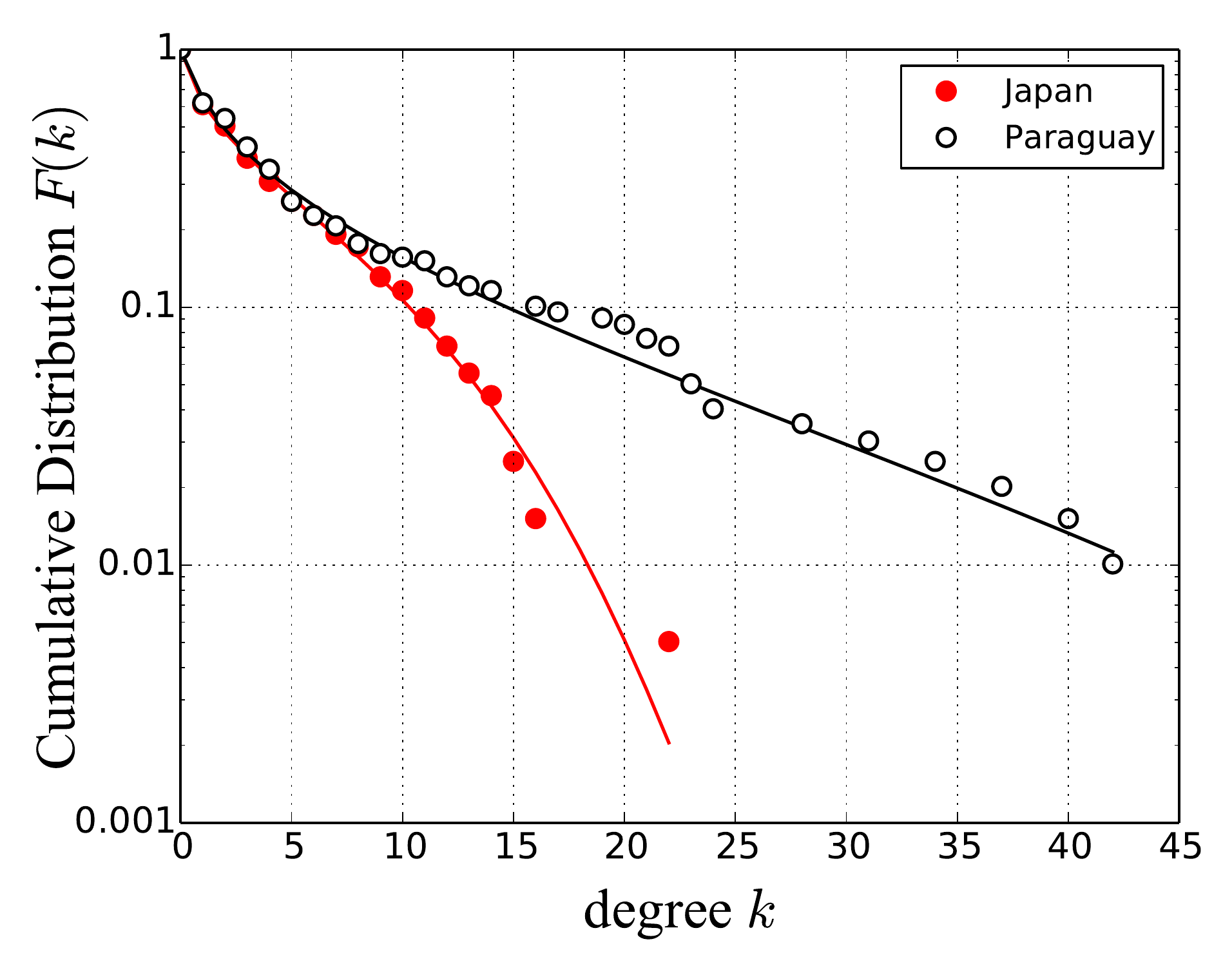}
		\caption*{(i)}
	\end{minipage}
	\begin{minipage}{.5\textwidth}
		\centering
		\includegraphics[width=6.5truecm]{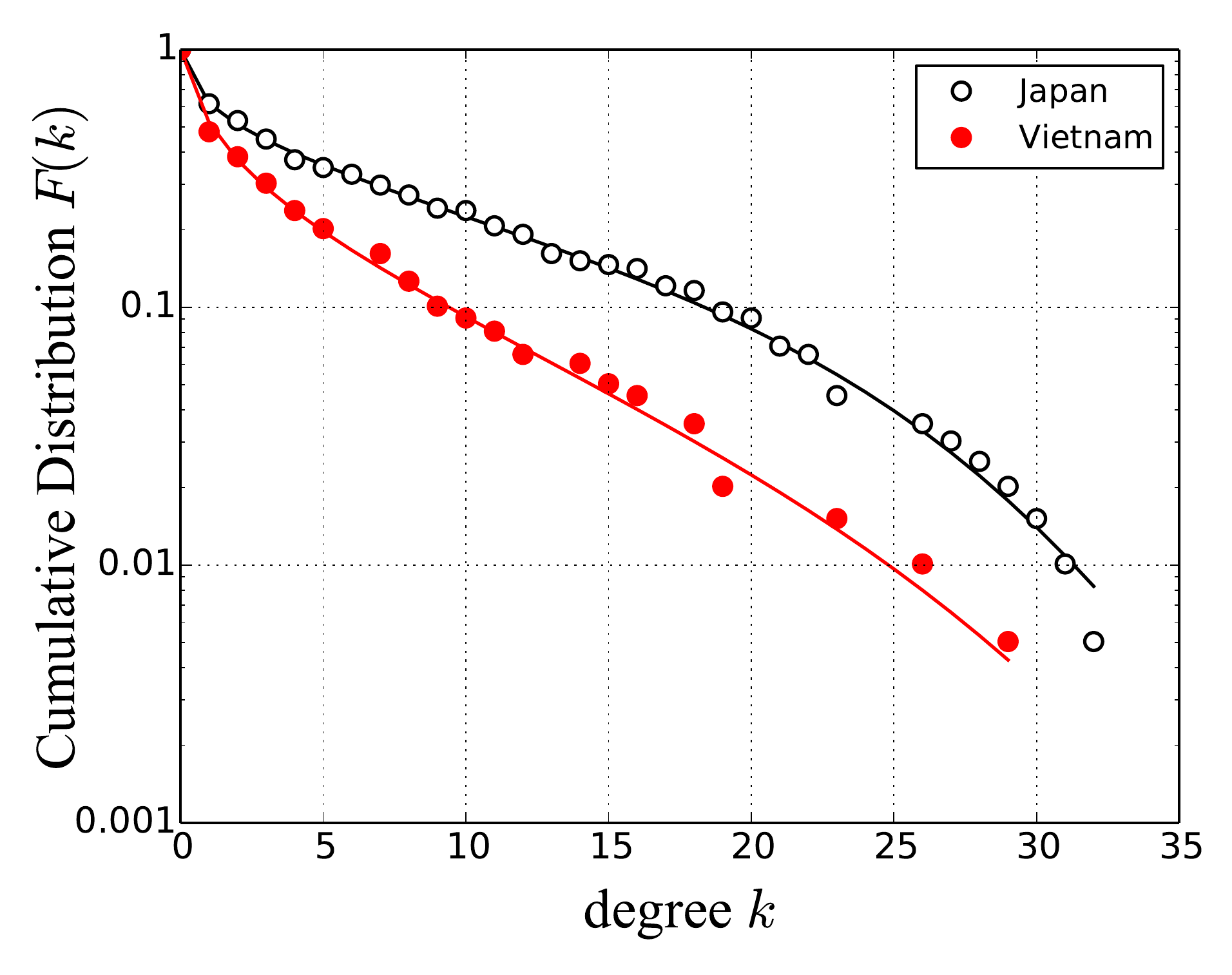}
		\caption*{(ii)}
	\end{minipage}
\end{figure}

\vspace*{-0.7truecm}

\begin{figure}[H]
	\begin{minipage}{.5\textwidth}
		\centering
		\includegraphics[width=6.5truecm]{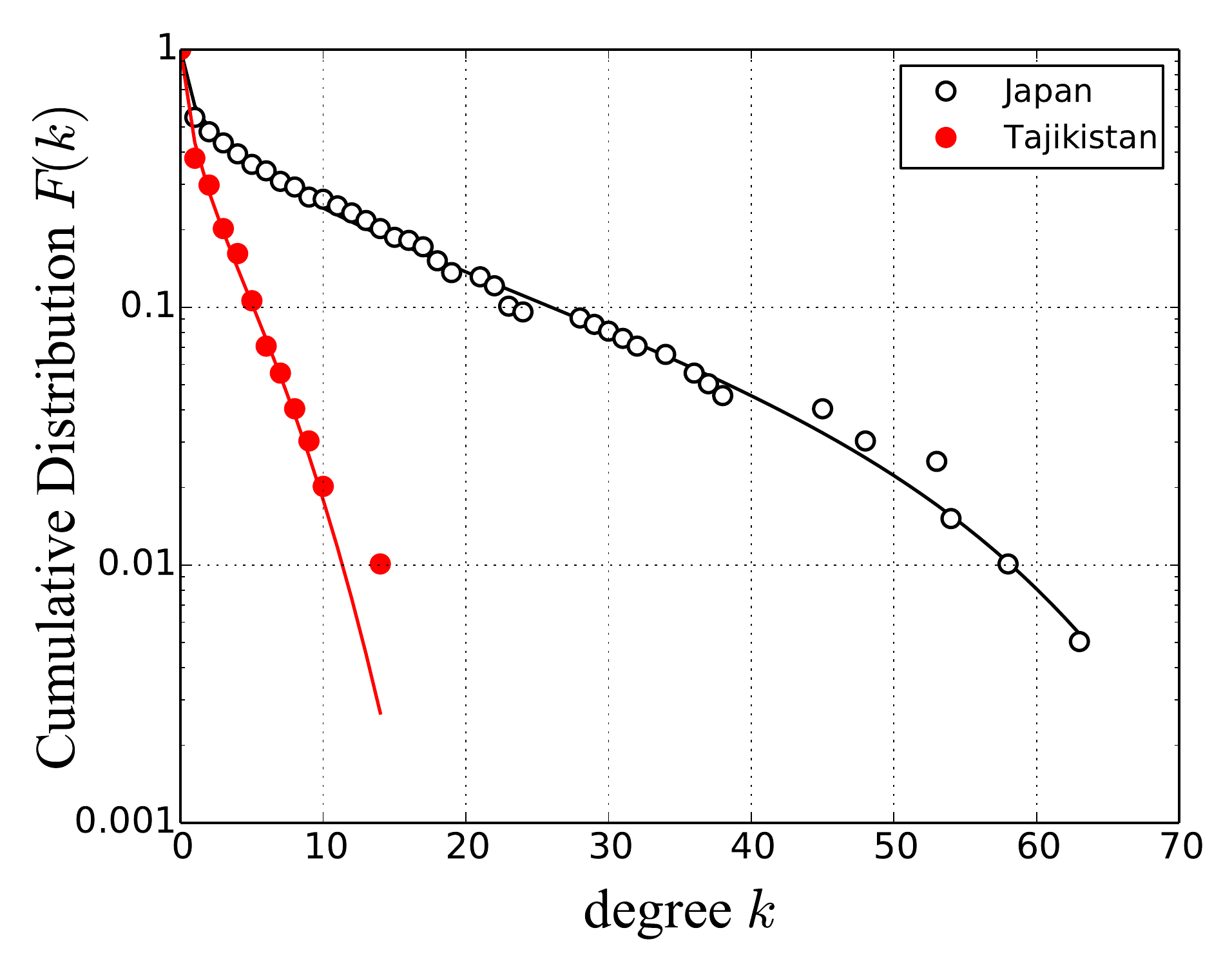}
		\caption*{(iii)}
	\end{minipage}
	\begin{minipage}{.5\textwidth}
		\centering
		\includegraphics[width=6.5truecm]{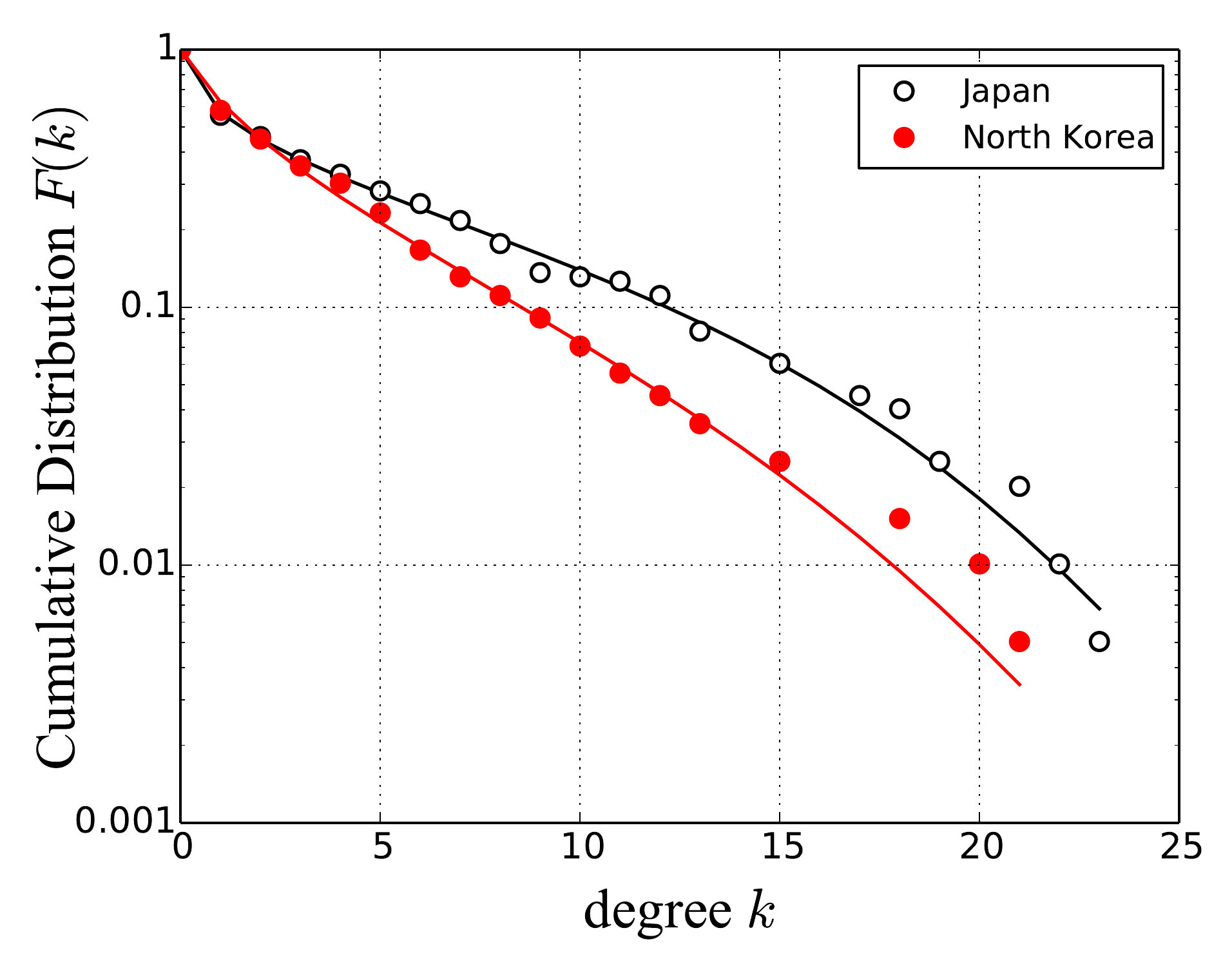}
		\caption*{(iv)}
	\end{minipage}
\end{figure}

\vspace*{-0.7truecm}

\begin{figure}[H]
	\begin{minipage}{.5\textwidth}
		\centering
		\includegraphics[width=6.5truecm]{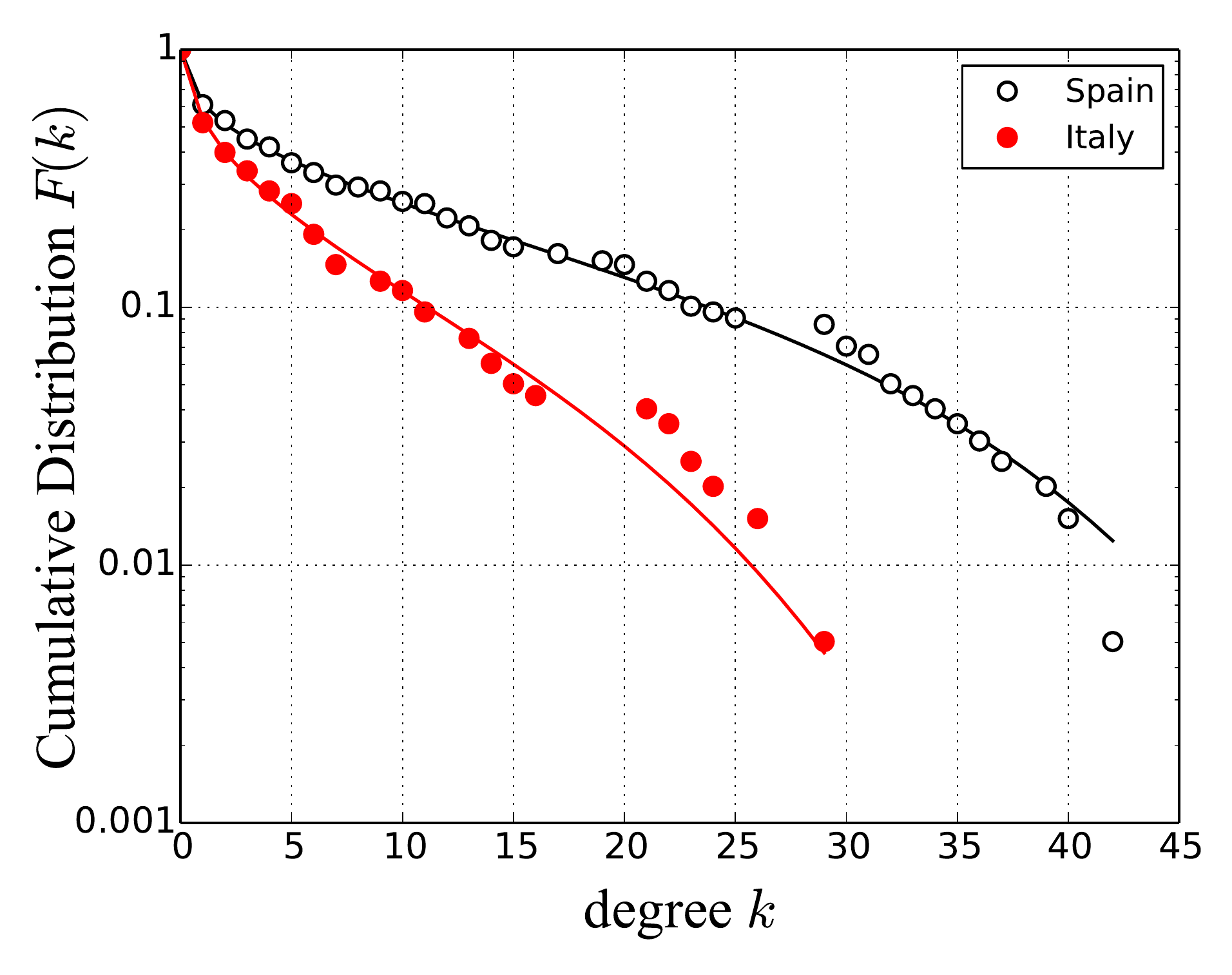}
		\caption*{(v)}
	\end{minipage}
	\begin{minipage}{.5\textwidth}
		\centering
		\includegraphics[width=6.5truecm]{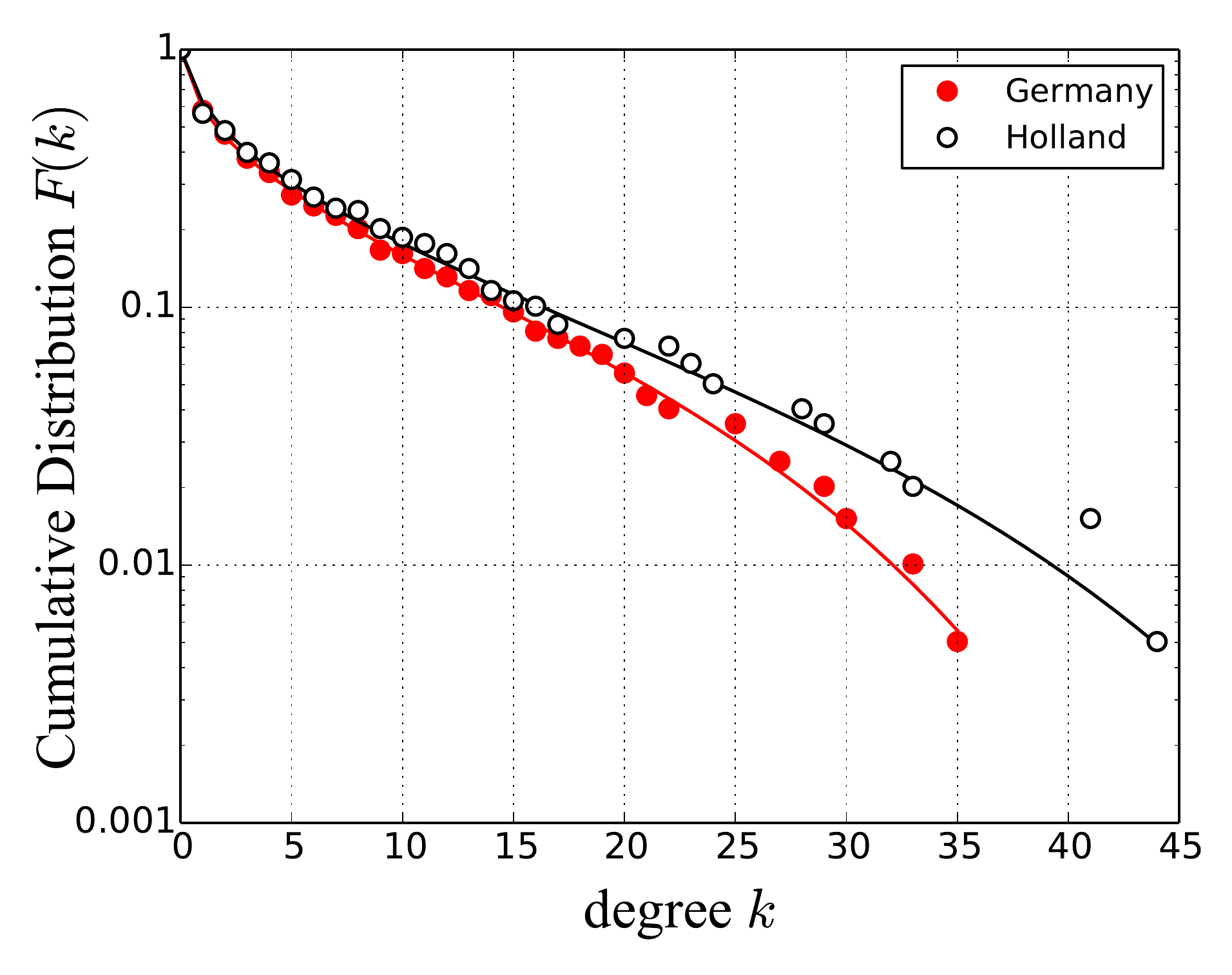}
		\caption*{(vi)}
	\end{minipage}
\end{figure}

\vspace*{-0.7truecm}

\begin{figure}[H]
	\begin{minipage}{.5\textwidth}
		\centering
		\includegraphics[width=6.5truecm]{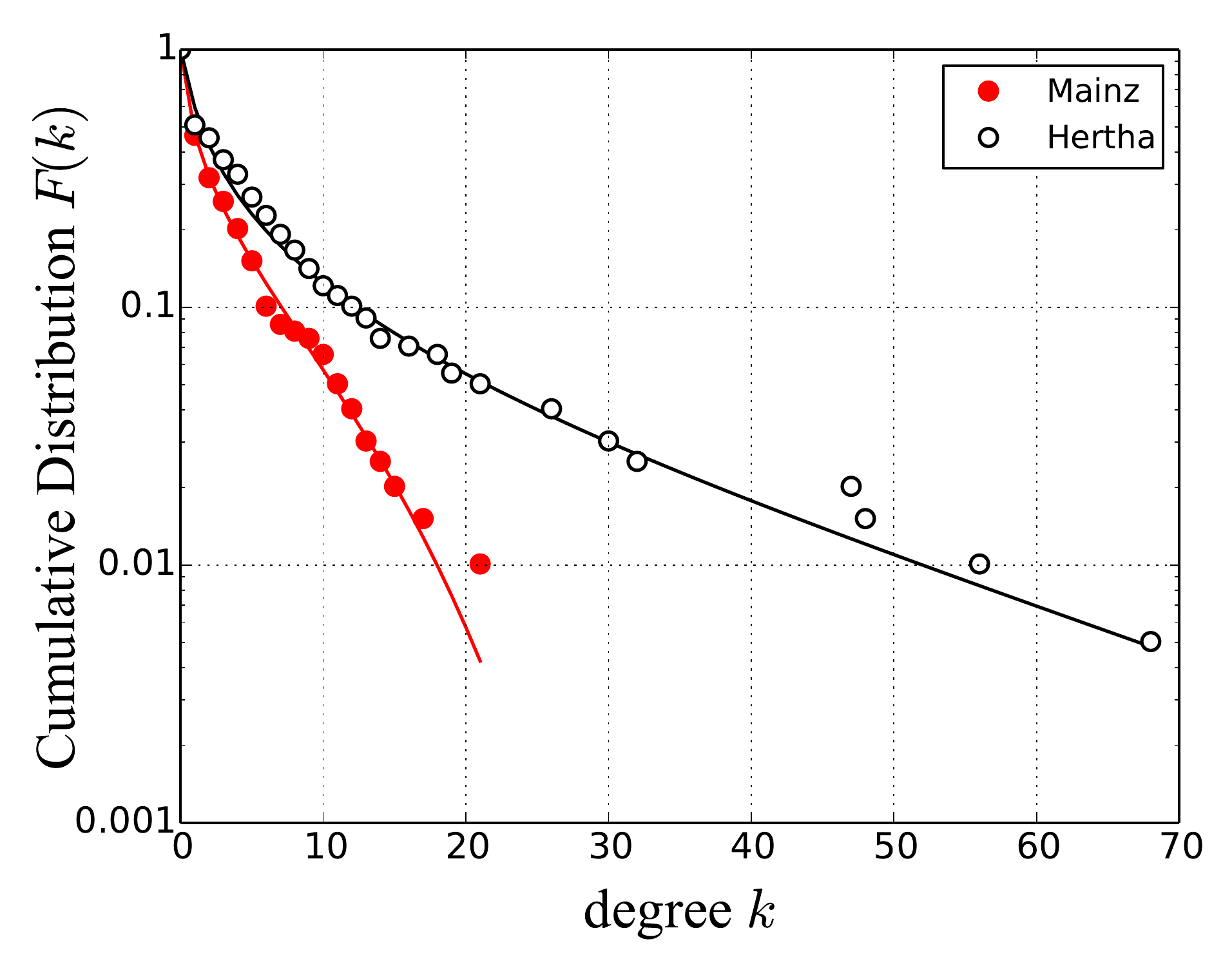}
		\caption*{(vii)}
	\end{minipage}
	\begin{minipage}{.5\textwidth}
		\centering
		\includegraphics[width=6.5truecm]{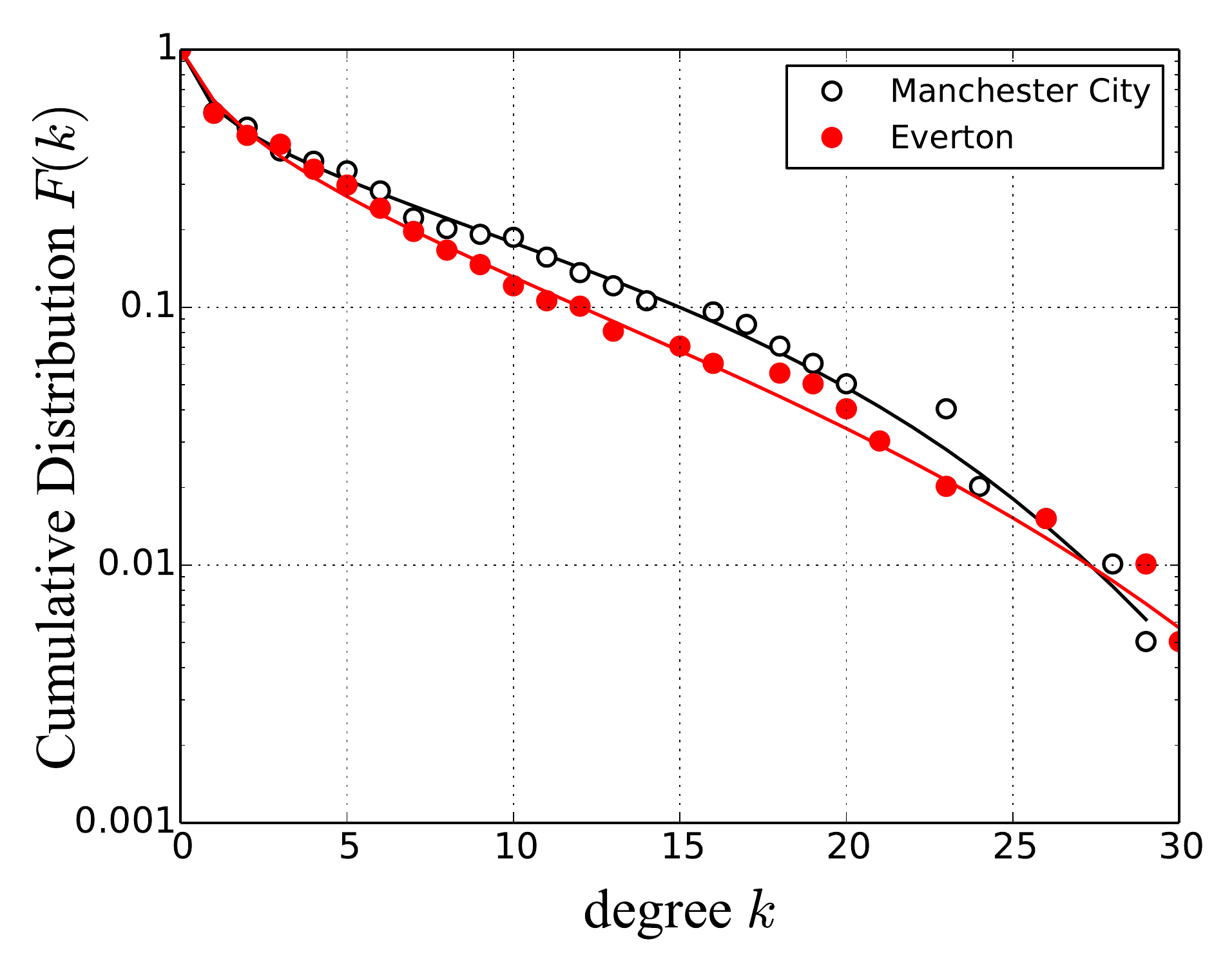}
		\caption*{(viii)}
	\end{minipage}
\end{figure}

\begin{table}[H]
  \centering
  \caption{The values of $ \mu $ and $ m $ for fittings of each network.}
  \vspace*{-0.3cm} 
    \begin{tabular}{ccccccc}
    \toprule
              Game            & Team              & $\mu$ & $ m $  \\ \toprule
	\multirow{2}[2]{*}{(i)}   & Japan             & 3.27  & 1.26    \\ 
                              & Paraguay          & 4.62  & 2.25   \\ \midrule   
    \multirow{2}[2]{*}{(ii)}  & Japan             & 3.92  & 1.10   \\ 
                              & Vietnam           & 4.70  & 1.80   \\ \midrule
    \multirow{2}[2]{*}{(iii)} & Japan             & 4.90  & 1.40   \\
                              & Tajikistan        & 4.20  & 1.50   \\ \midrule
    \multirow{2}[2]{*}{(iv)}  & Japan             & 3.97  & 1.14   \\
                              & North Korea       & 3.40  & 1.80   \\ \midrule
    \multirow{2}[2]{*}{(v)}   & Spain             & 4.36  & 1.17   \\
                              & Italy             & 4.57  & 1.53   \\ \midrule
    \multirow{2}[2]{*}{(vi)}  & Germany           & 4.35  & 1.52   \\
                              & Holland           & 4.40  & 1.70   \\ \midrule
    \multirow{2}[2]{*}{(vii)} & Mainz             & 4.67  & 1.71   \\
                              & Hertha            & 5.80  & 3.00   \\ \midrule
   	\multirow{2}[2]{*}{(viii)}& Manchester City   & 4.00  & 1.20   \\
                              & Everton           & 3.89  & 1.80   \\ \midrule
    \multirow{2}[2]{*}{(ix)}  & Manchester United & 4.00  & 2.00   \\
                              & Tottenham         & 3.80  & 2.40   \\ 
    \bottomrule
    \end{tabular}%
  \label{tb:params_fit}%
\end{table}%

\section{Discussion}
\begin{figure}[H]
	\centering
	\includegraphics[width=8truecm]{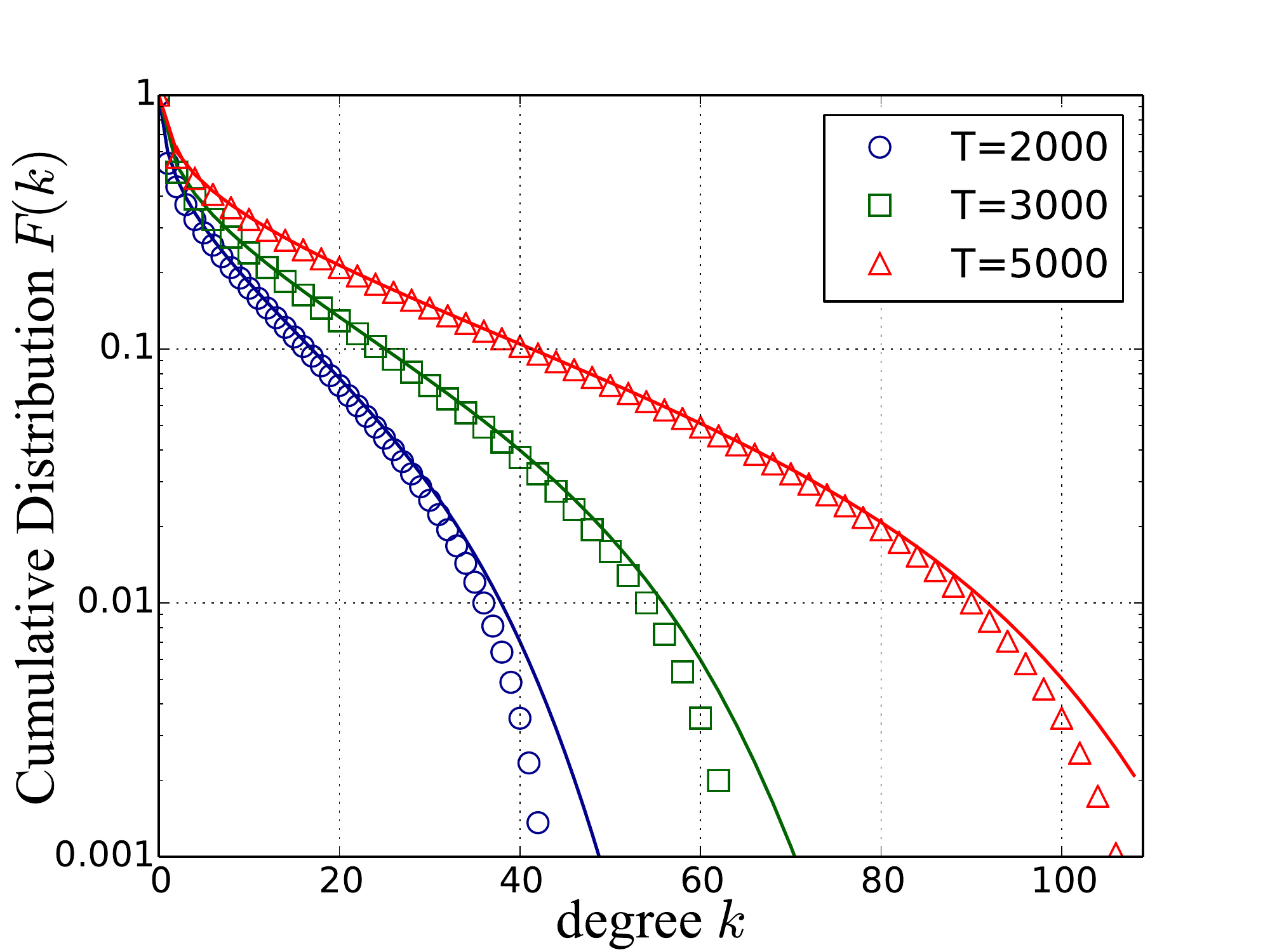}
	\caption{Comparison of cumulative degree distribution given by Eq. \eqref{eq:fk_real} (the solid curves) and Eq. \eqref{eq:fk_real_sim} (the points), where $ N_{A}=198 $, $ \eta_{A}=\eta_{B}=0.5 $, $ \mu=4.5 $, and $ m=1.5 $.}
	\label{fig:blw_lw}
\end{figure}

\begin{figure}[H]
	\begin{minipage}{.5\textwidth}
		\centering
		\includegraphics[width=6.5truecm]{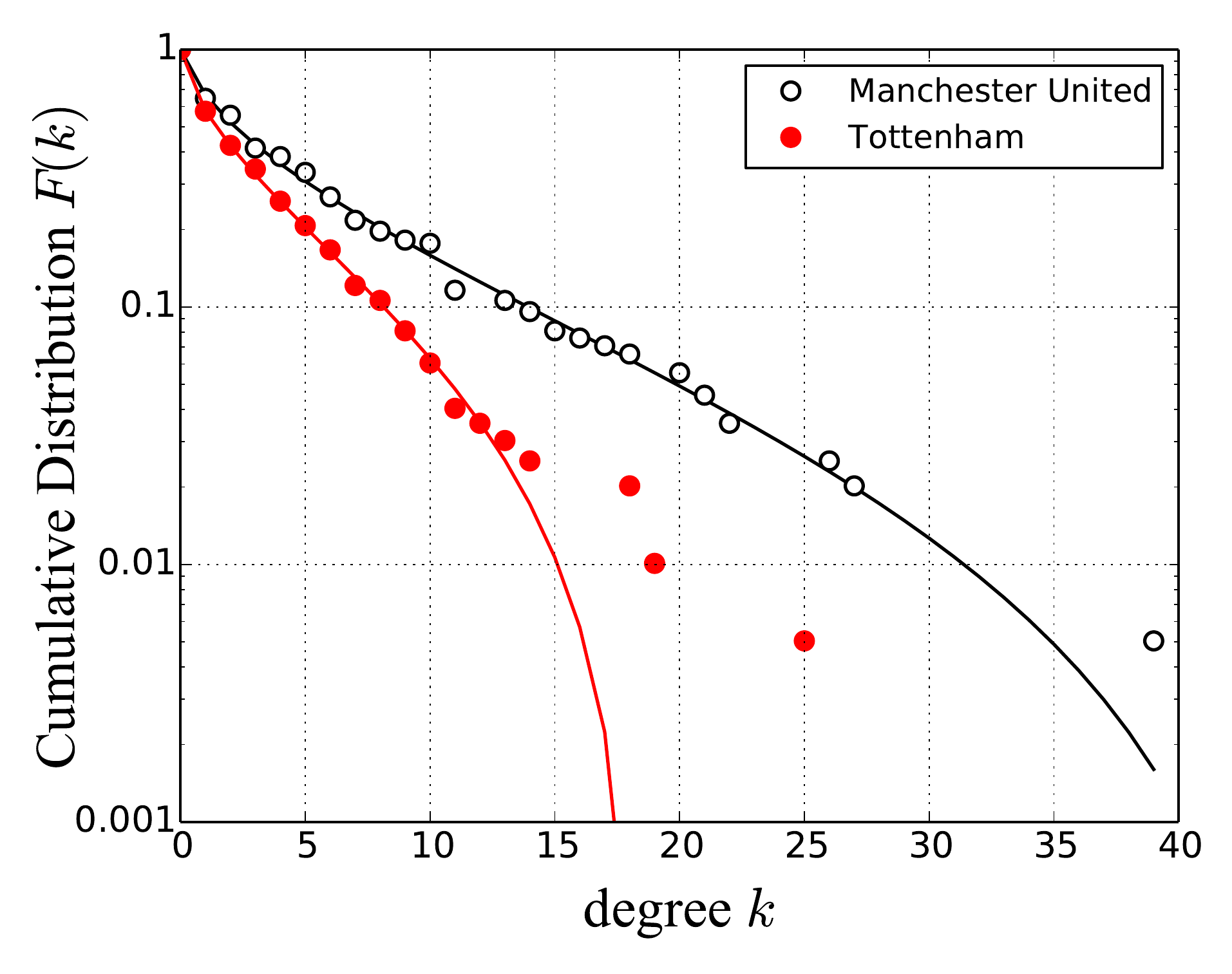}
		\caption*{(ix)}
	\end{minipage}
	\caption{Fitting for the cumulative distributions of real data by using Eq. \eqref{eq:fk_real_sim} (the solid curves).}
	\label{fig:degree2}
\end{figure}

Regarding Eq. \eqref{eq:fk_real}, a more simplified expression can be derived.
Since $ k/(2T) \lesssim 0.01 $ from the data in Table \ref{tb:params_data} and Fig. \ref{fig:degree}, the following approximation of Eq. \eqref{eq:fk_real} holds:
\begin{align}
	f_{A}(k) &\simeq \frac{m}{ \mu}\left(  \frac{\log\left( \bar{k}_{\textrm{max}}/k\right)}{\mu} \right)^{m-1} \frac{1}{k} \ 
						\exp \left[ -\left( \frac{\log\left( \bar{k}_{\textrm{max}}/k\right)}{\mu} \right)^{m} \right] ,
	\label{eq:fk_real_sim}
\end{align}
(see Appendix for detail discussion).
Here, $ \bar{k}_{\textrm{max}} $ is the expected value of the maximum degree given by
\begin{equation}
	\bar{k}_{\textrm{max}} = \frac{2T}{Z_{A}} \frac{\eta_{A}(1-\eta_{B})}{2-\eta_{A}-\eta_{B}}.
	\label{eq:kmax}
\end{equation}
In Fig. \ref{fig:blw_lw}, we compare the two functions \eqref{eq:fk_real} and \eqref{eq:fk_real_sim}.
It is found that the two functions take almost the same values except a high-degree part.
Such a mismatch in the high-degree part is also found in the results of fittings of the real data by Eq. \eqref{eq:fk_real_sim}.
However, if we also regard the maximum degree as another fitting parameter represented by $ k^{\textrm{(fit)}}_{\textrm{max}} $, all degree distributions are fitted well by Eq. \eqref{eq:fk_real_sim} as shown in Fig. \ref{fig:degree2}.
The values of $ k^{\textrm{(fit)}}_{\textrm{max}} $ and the maximum degree of real data $ k^{\textrm{(real)}}_{\textrm{max}} $ are summarised in Table. \ref{tb:kmax}.
Although $ k^{\textrm{(fit)}}_{\textrm{max}} $ and $ k^{\textrm{(real)}}_{\textrm{max}} $ are different in some teams such as Paraguay in the game (i) and Hertha in the game (vii), we conclude that Eq. \eqref{eq:fk_real_sim} roughly describes the degree distributions of ball-passing networks.

Eq. \eqref{eq:fk_real_sim} is the Weibull distribution where we set $ y =  \log\left( \bar{k}_{\textrm{max}}/k\right) $:
\begin{align}
	f_{A}(y) &= 	\frac{m}{ \mu}\left(  \frac{y}{\mu} \right)^{m-1} 
						\exp \left[ -\left( \frac{y}{\mu} \right)^{m} \right].  \nonumber
\end{align}
Here $ m $ and $ \mu $ correspond to the shape and scale parameters, respectively.
(Eq. \eqref{eq:fk_real_sim} has been used for the statistics in earthquake \cite{Huillet1999, Hasumi2009}, for instance.)
We note that Eq. \eqref{eq:fk_real_sim} becomes the power-law function $ f_{A}(k) \sim k^{\frac{1}{\mu}-1}$ when $ m=1 $, which corresponds to $ \lambda \to \infty $ in the truncated-gamma function [Eq. \eqref{eq:tg}].
In the real data, some teams fulfill this condition; Japan in the game (i) , (ii) and (iv), Spain in (v) and Manchester City in (viii) have the value of $ m $ close to 1 (see Table \ref{tb:params_fit}).
This is why the truncated-gamma distribution also fits to the real data.

\clearpage
\begin{figure}[H]
	\begin{minipage}{.5\textwidth}
		\centering
		\includegraphics[width=6.5truecm]{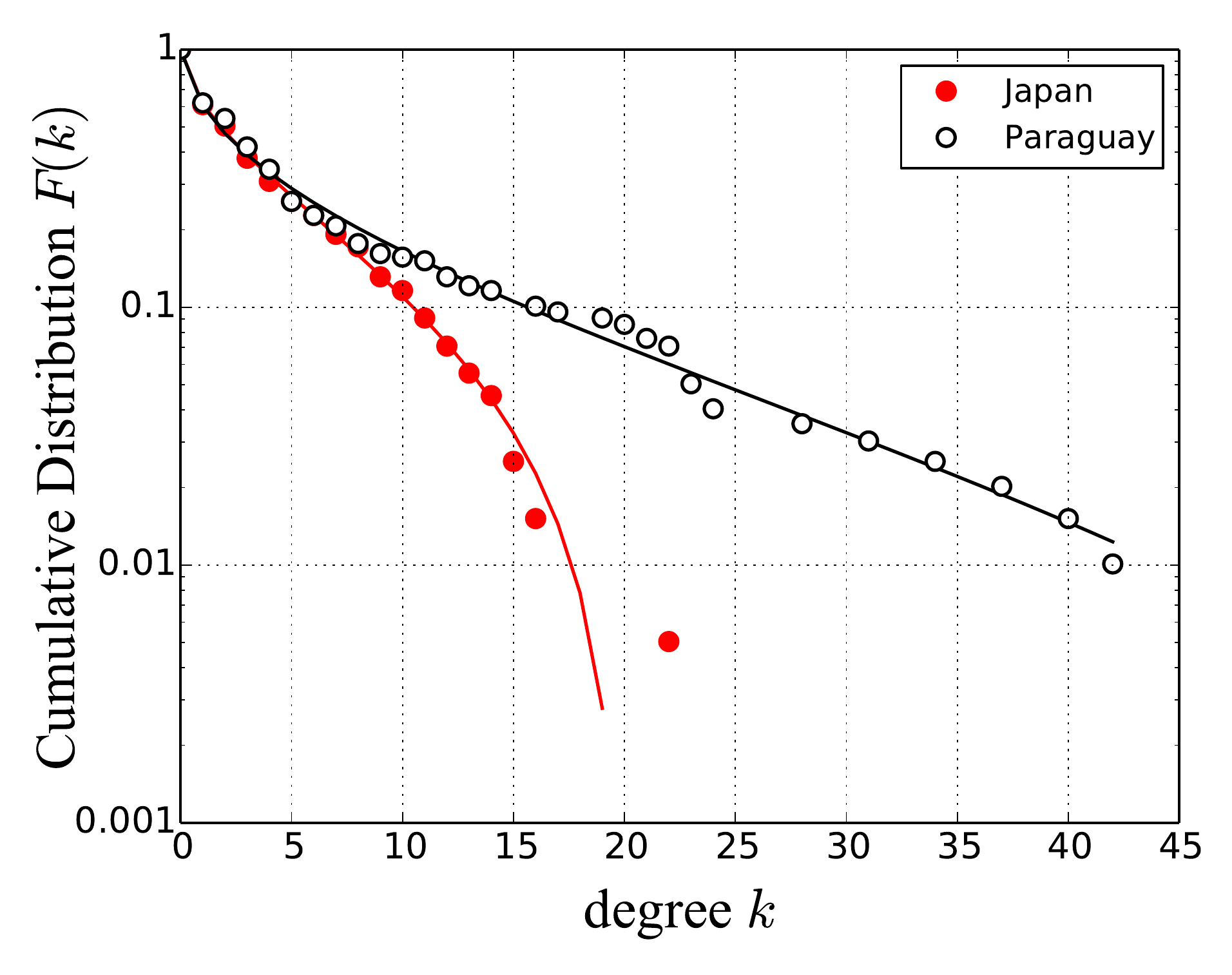}
		\caption*{(i)}
	\end{minipage}
	\begin{minipage}{.5\textwidth}
		\centering
		\includegraphics[width=6.5truecm]{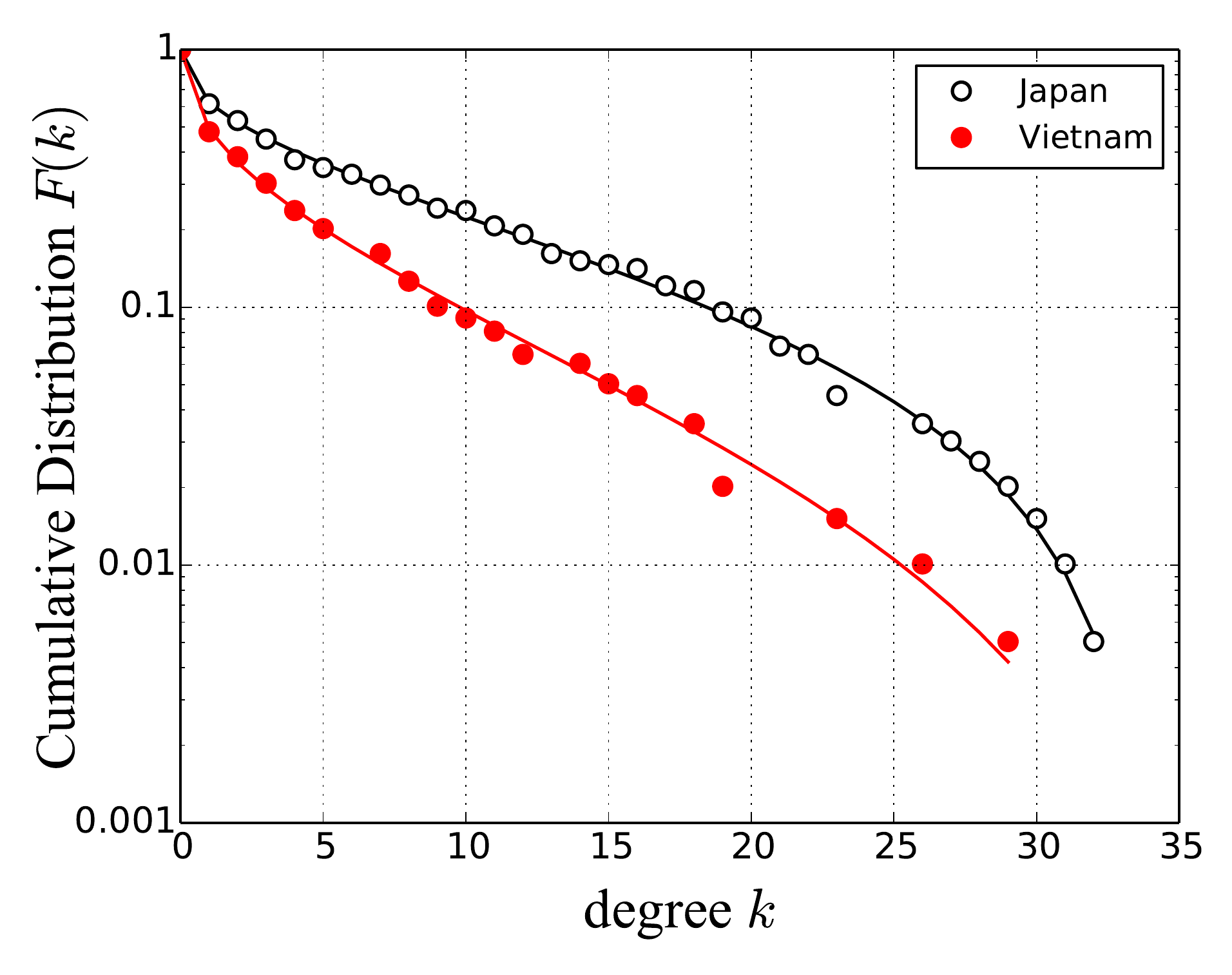}
		\caption*{(ii)}
	\end{minipage}
\end{figure}

\vspace*{-0.7truecm}

\begin{figure}[H]
	\begin{minipage}{.5\textwidth}
		\centering
		\includegraphics[width=6.5truecm]{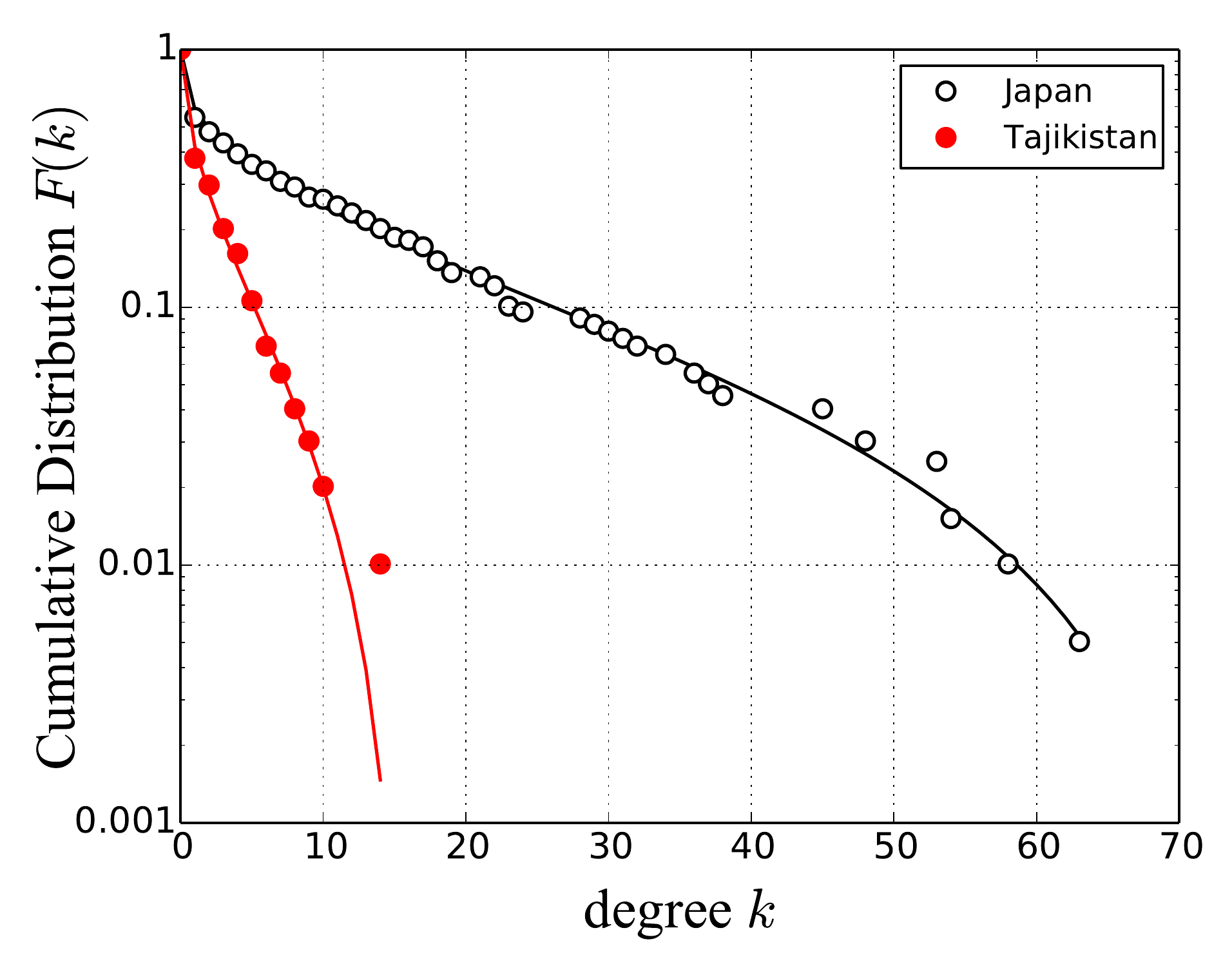}
		\caption*{(iii)}
	\end{minipage}
	\begin{minipage}{.5\textwidth}
		\centering
		\includegraphics[width=6.5truecm]{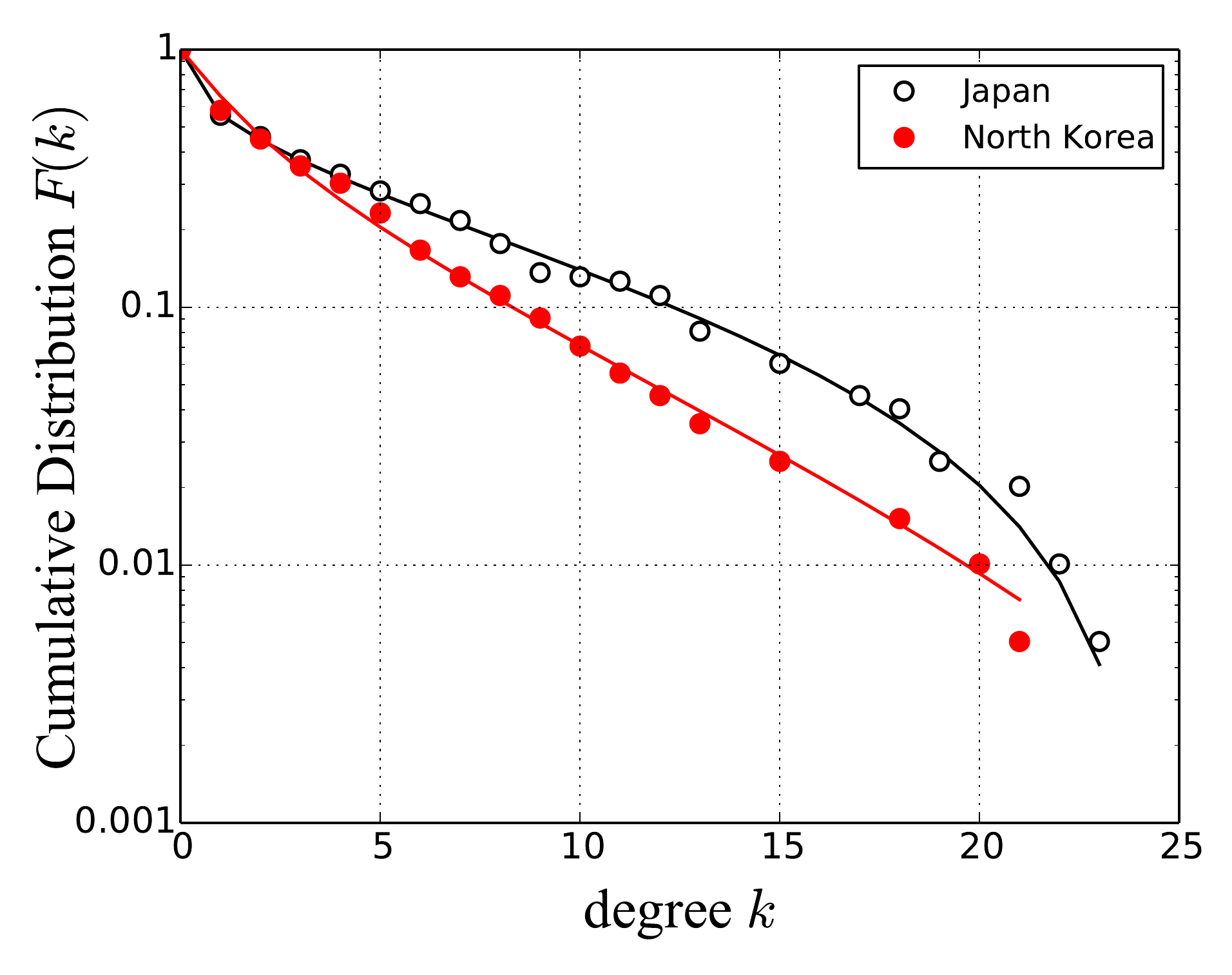}
		\caption*{(iv)}
	\end{minipage}
\end{figure}

\vspace*{-0.7truecm}

\begin{figure}[H]
	\begin{minipage}{.5\textwidth}
		\centering
		\includegraphics[width=6.5truecm]{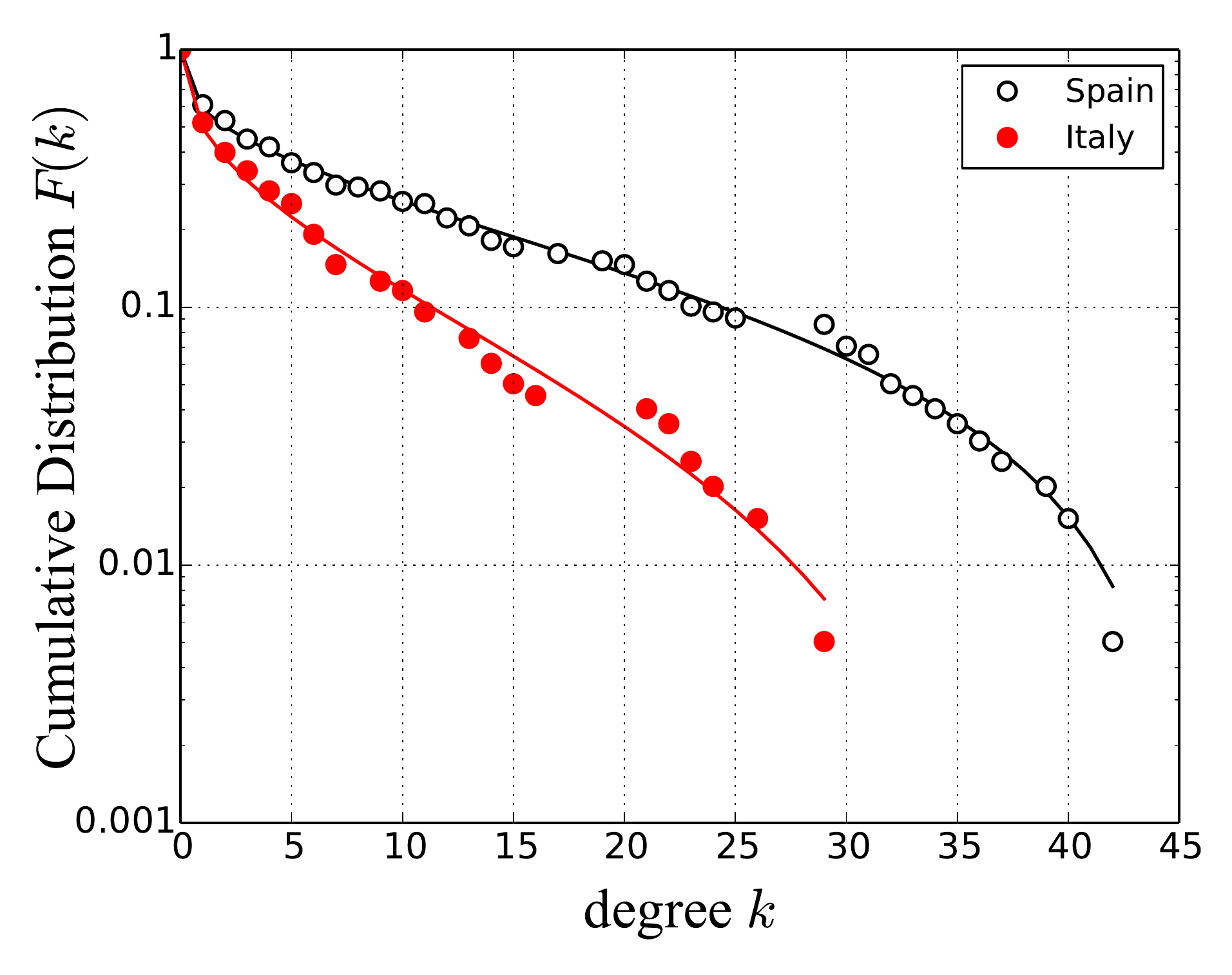}
		\caption*{(v)}
	\end{minipage}
	\begin{minipage}{.5\textwidth}
		\centering
		\includegraphics[width=6.5truecm]{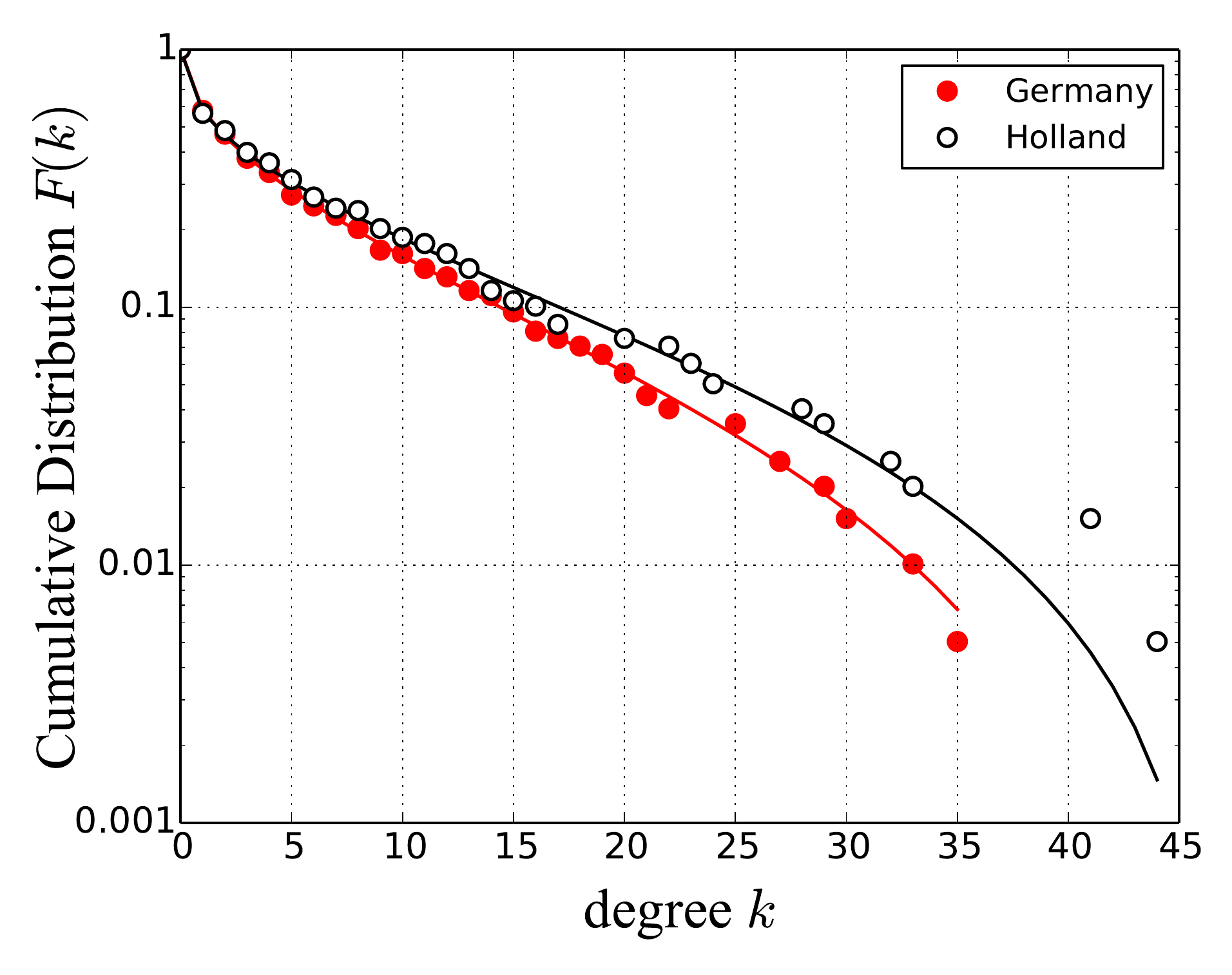}
		\caption*{(vi)}
	\end{minipage}
\end{figure}

\vspace*{-0.7truecm}

\begin{figure}[H]
	\begin{minipage}{.5\textwidth}
		\centering
		\includegraphics[width=6.5truecm]{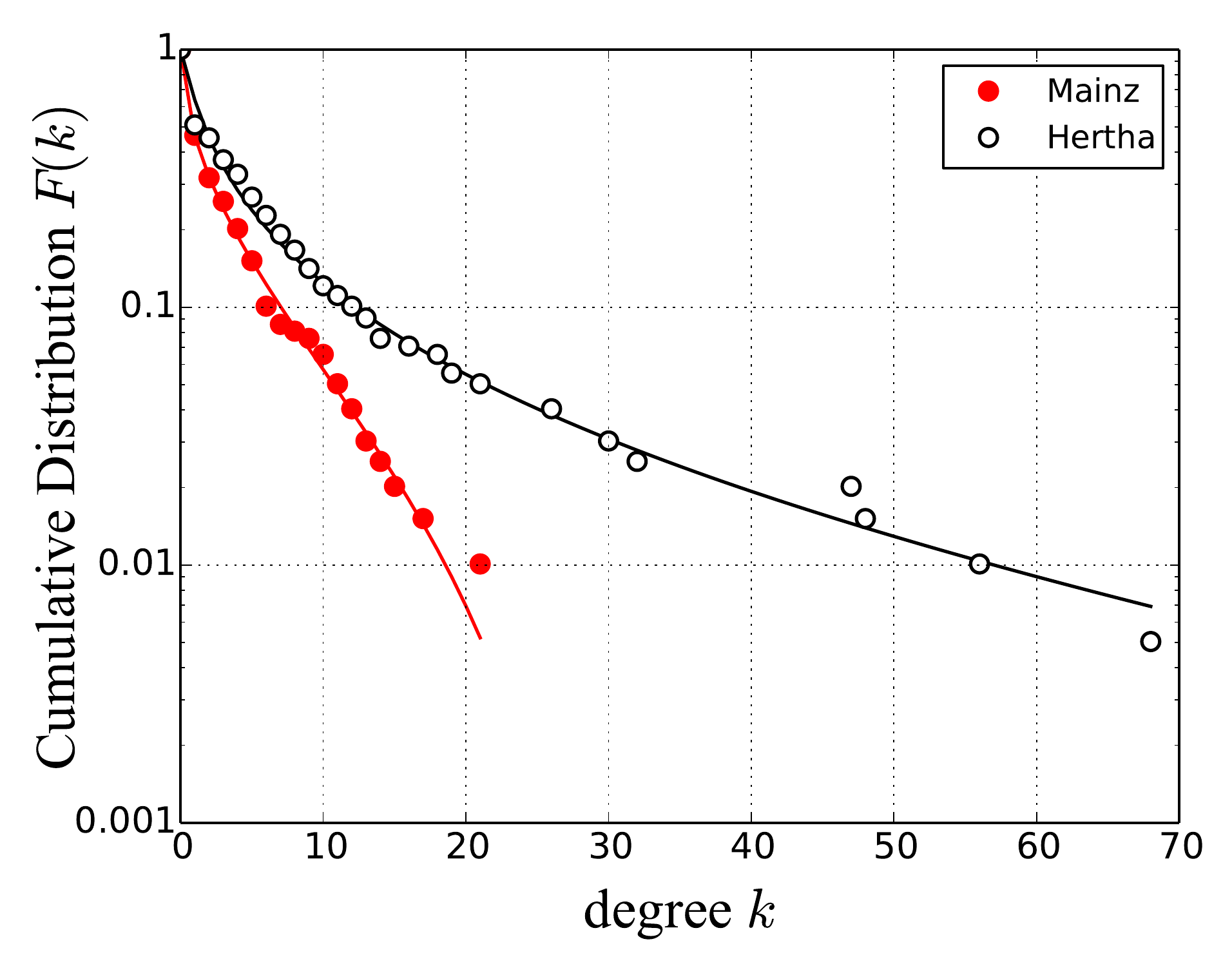}
		\caption*{(vii)}
	\end{minipage}
	\begin{minipage}{.5\textwidth}
		\centering
		\includegraphics[width=6.5truecm]{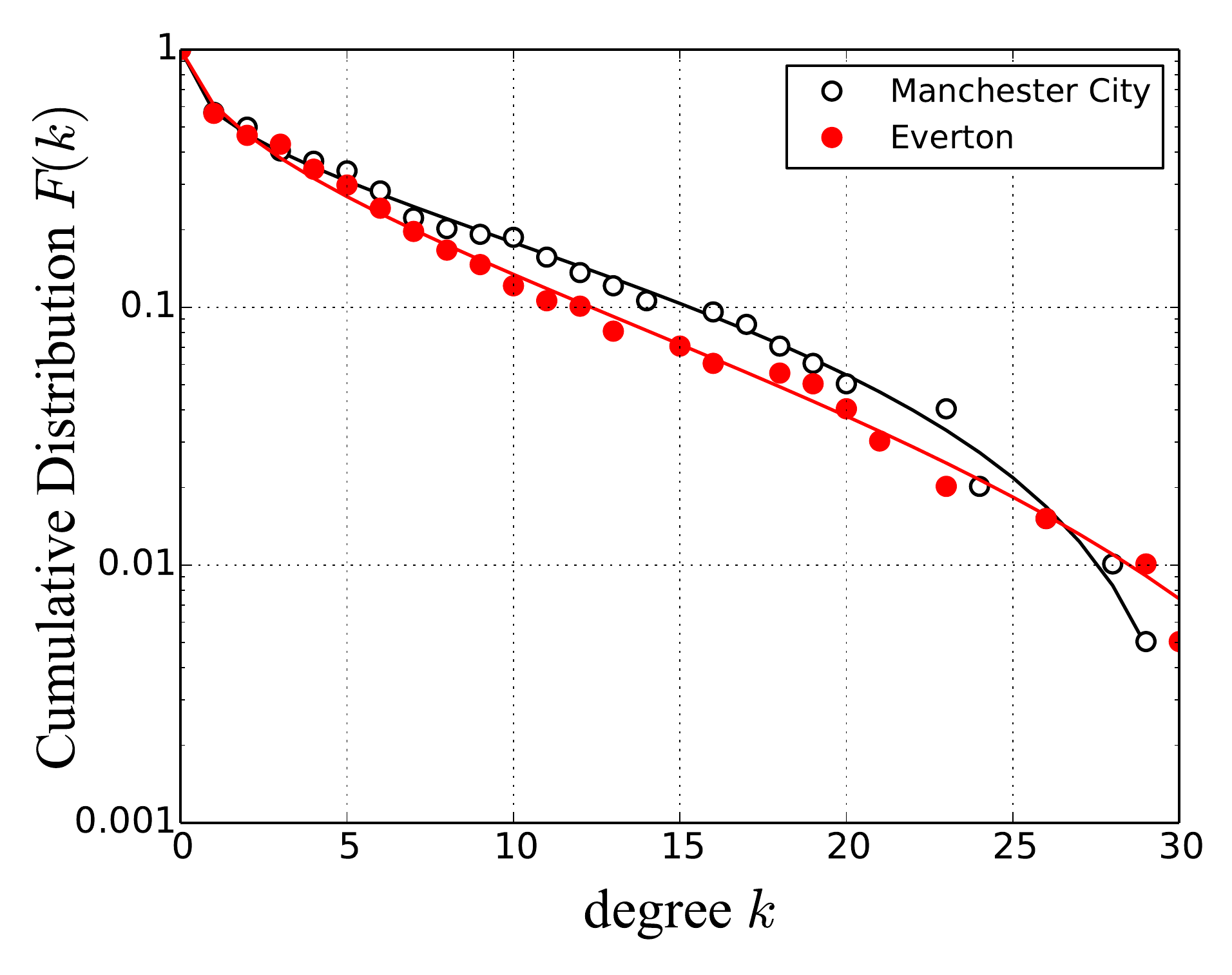}
		\caption*{(viii)}
	\end{minipage}
\end{figure}

\begin{table}[H]
  \centering
  \caption{The values of maximum degree.}
  \vspace*{-0.3cm} 
    \begin{tabular}{ccccccccc}
    \toprule
              Game            & Team              & $k^{(\textrm{fit})}_{\textrm{max}}$ & $ k^{(\textrm{real})}_{\textrm{max}} $ \\ \toprule
	\multirow{2}[2]{*}{(i)}   & Japan             & 20.0 & 22.0 \\ 
                              & Paraguay          & 75.0 & 42.0 \\ \midrule   
    \multirow{2}[2]{*}{(ii)}  & Japan             & 33.9 & 32.0 \\ 
                              & Vietnam           & 36.5 & 29.0 \\ \midrule
    \multirow{2}[2]{*}{(iii)} & Japan             & 72.0 & 63.0 \\
                              & Tajikistan        & 15.5 & 14.0 \\ \midrule
    \multirow{2}[2]{*}{(iv)}  & Japan             & 24.4 & 23.0 \\
                              & North Korea       & 35.0 & 21.0 \\ \midrule
    \multirow{2}[2]{*}{(v)}   & Spain             & 45.0 & 42.0 \\
                              & Italy             & 37.0 & 29.0 \\ \midrule
    \multirow{2}[2]{*}{(vi)}  & Germany           & 44.0 & 35.0 \\
                              & Holland           & 47.0 & 44.0 \\ \midrule
    \multirow{2}[2]{*}{(vii)} & Mainz             & 29.1 & 21.0 \\
                              & Hertha            & 500.0 & 68.0 \\ \midrule
   	\multirow{2}[2]{*}{(viii)}& Manchester City   & 31.1 & 29.0 \\
                              & Everton           & 40.5 & 30.0 \\ \midrule
    \multirow{2}[2]{*}{(ix)}  & Manchester United & 45.0 & 39.0 \\
                              & Tottenham         & 18.4 & 25.0 \\ 
    \bottomrule
    \end{tabular}%
  \label{tb:kmax}%
\end{table}%

\begin{figure}[H]
	\begin{minipage}{.5\textwidth}
		\centering
		\includegraphics[width=7.5truecm]{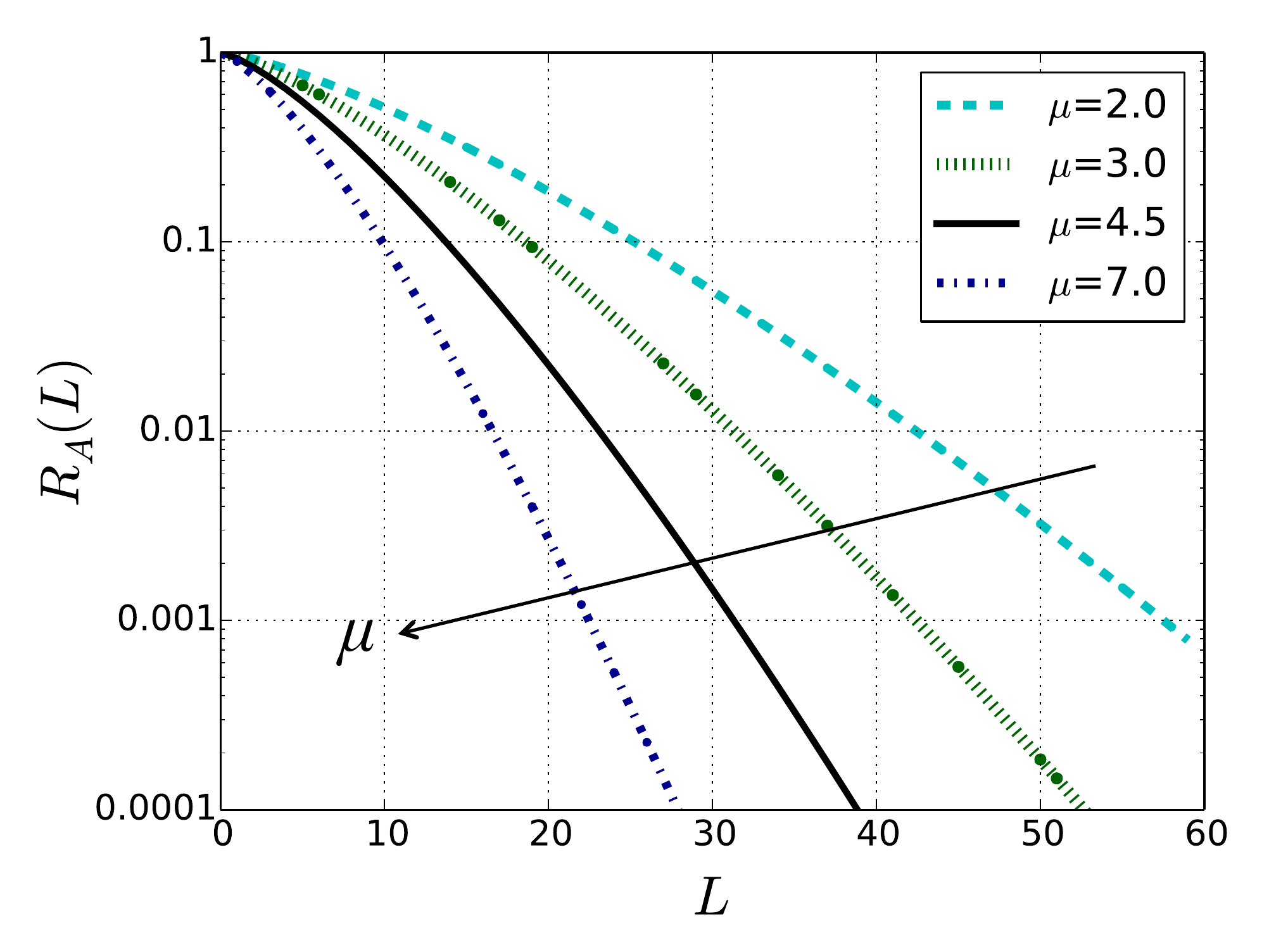}
		\caption*{(i)}
	\end{minipage}
	\begin{minipage}{.5\textwidth}
		\centering
		\includegraphics[width=7.5truecm]{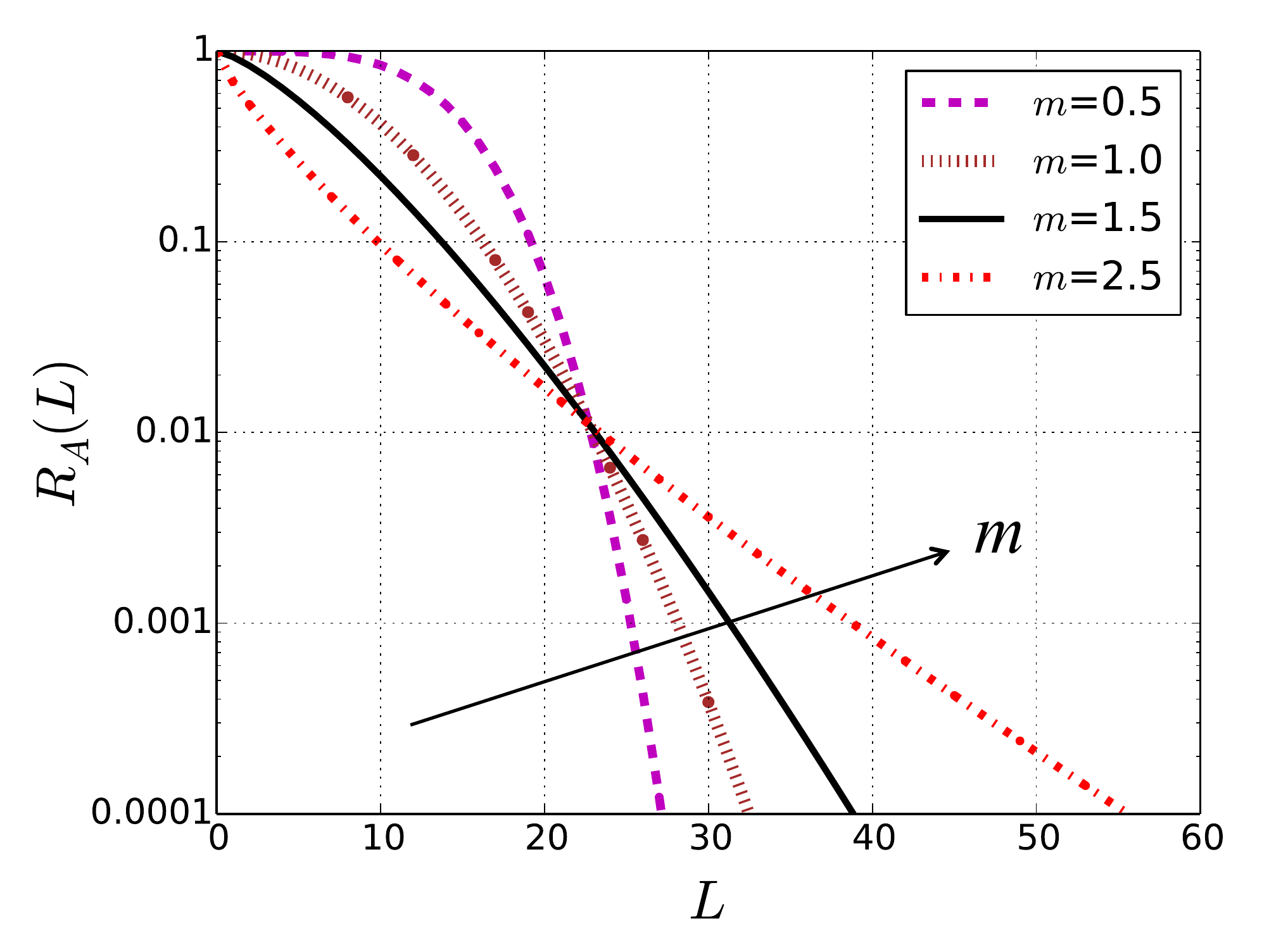}
		\caption*{(ii)}
	\end{minipage}
\caption{Change of the shape of $ R_{A}(L) $ with $ \mu $ and $ m $. (i) $ \mu $-dependence with $ m=1.5 $, (ii) $ m $-dependence with $ \mu=4.5 $.}
\label{fig:R}
\end{figure}

In the present study, we have derived the explicit expression of the degree distribution [Eq. \eqref{eq:fk_real}] from the extended Markov-chain model.
In contrast, our previous study \cite{Narizuka2014} heuristically adopted the truncated-gamma distribution for fitting.
Although both functions fit to the real data well, Eq. \eqref{eq:fk_real} has the following three advantages over the previous one.
First, Eq. \eqref{eq:fk_real} is derived from the model based on the simple passing process.
The key point of our model is that $ P_{i\to j} $ depends only on $ \eta_{A} $, $ \eta_{B} $, and  $ R(L_{j}) $.
This is a considerable simplification of the real football games.
Nevertheless, Fig. \ref{fig:degree} shows that such simplification even preserves essential features of the actual degree distributions.
The second point is the number of control parameters.
The truncated-gamma distribution [Eq. \eqref{eq:tg}] needs to control the maximum degree $ k_{\textrm{max}} $ to reproduce the real data.
On the other hand, Eq. \eqref{eq:fk_real} does not contain the control parameter corresponding to the maximum degree since the domain of $ k $ of Eq. \eqref{eq:fk_real} is given as $ 0 \leq k < \infty $.
Thus, Eq. \eqref{eq:fk_real} can reproduce the real degree distributions with the only two parameters $ \mu $ and $ m $, which are fewer than the truncated-gamma distribution.
Third, $ \mu $ and $ m $ have the clear physical meaning as follows.
As shown in Sec. 3, $ \omega $ is calculated as  $ \omega = 22.72\pm 1.34 $ in the case where $ \Delta = 25 $, so that $ \mu \simeq (22.7/\beta)^{(2/m)} $.
Namely, both $ \mu $ and $ m $ are determined from the function $ R_{A}(L) $.
We illustrate the dependence of $ R_{A}(L) $ on $ \mu $ and $ m $ in Fig. \ref{fig:R}.
This figure shows that the variance of $ R_{A}(L) $ decreases with $ \mu $, and increases with $ m $.
Since $ R_{A}(L) $ represents the existence probability of a player dependent on the distance $ L $ from its home position, $ \mu $ and $ m $ control the typical moving range of the player around its home position.
In this sense, $ \mu $ and $ m $ reflect the strategy or playing style of a team.
However, we emphasize that the degree distribution does not change greatly by such differences of each game.
We expect that $ R_{A}(L) $ is determined directly by the analysis based on the more detailed positional data of each player.

From the viewpoint of complex network, our model is related to the fitness model \cite{Caldarelli2002, Boguna2003}.
The difference of the two models is as follows.
In the original fitness model, the algorithm of the network creation is static, and that an undirected network with un-weighted edges is generated.
On the other hand, our model contains the temporal features, i.e., the time series of the passes, and generates a directed network with weighted edges.
Hence, our model is not only the extension of the original fitness model, but also a kind of a temporal network \cite{Holme2012}.
The ball-passing networks in football game and the networks generated from our model seem to be similar to the co-occurrence network for human language \cite{Ferrer2001, Dorogovtsev2001} in that the following method is in common use; a node represents a word, and an directed edge connects the two adjacent words in the same sentence.
In creation of sentences, this type of co-occurrence network reflects the process of consecutive choice of words and its time series. 
Moreover, in such a creation process, each word is considered to be chosen according to its importance which is associated with the fitness. 
Therefore the co-occurrence network is expected to modelled by our Markov-chain model.
We believe that our model makes a theoretical contribution to network science, as well as the practical analysis of football games.

\section{Conclusion}
In the present paper, we have extended the previous model describing the position-dependent ball-passing network in football games.
In the extended model, we have taken the effect of the opponent team into account, and assumed that the transition probability depends on the existence probability $ R(L_{j}) $ of each node.
The explicit expression of the degree distribution (13) which has two parameters $\mu$ and $m$ are derived, and it can reproduce the real degree distributions quite well.
We have also found that $\mu $ and $ m $ are determined from the function $ R(L) $, and they characterize the moving range of a player in each team.
Furthermore, we have shown that Eq. \eqref{eq:fk_real} is simplified to the function \eqref{eq:fk_real_sim} when $ 2T \gg k $ is satisfied, and this simple function also approximates the real distribution well.
Although our model is simple, it incorporates the essential features of the formation of ball-passing networks in the actual football games.

\clearpage
\section*{Appendix: Derivation of Eq. \eqref{eq:fk_sim} from Eq. \eqref{eq:fk_2}}
Eq. \eqref{eq:fk_2} can be written as
\begin{align}
	 f_{A}(k) = -\binom{2T}{k} \int \mathrm{e}^{{\varphi(u)}}h(u) du,
	 \label{eq:int1}
\end{align}
where
\begin{align}
	\varphi(u) &= k \ln u + (2T-k)\ln(1-u) \nonumber, \\
	h(u) &= \rho_{A}\left[ R^{-1}_{A}\left(  Z_{A}'u \right) \right]
	       	  \frac{d}{du} \left[ R^{-1}_{A}\left(Z_{A}' u\right)\right] \nonumber.
\end{align}
$ \varphi(u) $ has the peak at $ u = k/2T $.
Expanding $ \varphi(u) $ to the second order near $ u = k/2T $, and substituting it into Eq. \eqref{eq:int1}, we get
\begin{align}
	f_{A}(k) \simeq -\binom{2T}{k} \exp\left[\varphi\left(\frac{k}{2T} \right)\right] 
	                   \int \exp\left[-\frac{4T^{3}}{k(2T-k)}\left(u - \frac{k}{2T} \right)^{2}\right] h(u) du.
	\label{eq:int2}
\end{align}
When $ 2T \gg k $, the integrand in Eq. \eqref{eq:int2} decreases rapidly without $ u=k/2T $, and it is allowed to expand the integration range to $ (-\infty, \infty) $.
Hence, we can approximate Eq. \eqref{eq:int2} as follows:
\begin{align}
	f_{A}(k) &\simeq -\binom{2T}{k} \exp\left[\varphi\left(\frac{k}{2T} \right)\right] 
	         \int_{-\infty}^{\infty} \exp\left[-\frac{4T^{3}}{k(2T-k)}\left(u-\frac{k}{2T} \right)^{2}\right] du \ 
	             h\left(\frac{k}{2T}\right) \nonumber \\
	  &= -\binom{2T}{k} \exp\left[\varphi\left(\frac{k}{2T} \right)\right] \sqrt{\frac{k(2T-k)\pi}{4T^{3}}} 
	          h\left(\frac{k}{2T}\right).
	\label{eq:int3}
\end{align}
Here, we apply Stirling's formula, $ t! \simeq \sqrt{2\pi t} \ t^{t} \mathrm{e}^{-t} $, to Eq. \eqref{eq:int3}, and use
\begin{align}
	 \exp\left[\varphi\left(\frac{k}{2T} \right)\right] = \frac{k^{k}(2T-k)^{2T-k}}{(2T)^{2T}}, \nonumber
\end{align}
we obtain Eq. \eqref{eq:fk_sim}:
\begin{align}
	f_{A}(k) &\simeq -\frac{1}{2T} h\left(\frac{k}{2T}\right)\nonumber \\
	  &= -\rho_{A}\left[ R^{-1}_{A}\left(  Z_{A}'k/2T \right) \right]
	 	 \frac{d}{dk} \left[ R^{-1}_{A}\left(Z_{A}' k/2T \right)\right] \nonumber.
\end{align}

\clearpage
\bibliography{./reference} 
\end{document}